\documentclass[a4paper]{amsart}

\usepackage{lineno}
\usepackage[hidelinks]{hyperref}

\usepackage{amsmath}
\usepackage{amssymb}
\usepackage{amsthm}
\usepackage{hyperref}
\usepackage{subfig}
\usepackage{graphicx}
\usepackage{graphbox} 
\usepackage[left=3cm, right=3cm, bottom=3cm, top=3cm]{geometry}

\usepackage{float}
\usepackage{xcolor}
\hypersetup{
    colorlinks,
    linkcolor={red!50!black},
    citecolor={blue!50!black},
    urlcolor={blue!80!black}
}
\usepackage[foot]{amsaddr}

\theoremstyle{plain} 

\usepackage{mathtools}
\usepackage{pgfplots}
\pgfplotsset{/pgf/number format/use comma,compat=newest}

\renewcommand\epsilon{\varepsilon}
\newcommand{\R}{\mathbb{R}}

\newcommand{\vect}[1]{\boldsymbol{#1}}

\newtheorem*{remark*}{\bf Remark}
\begin{document}
\title{Rods coiling about a rigid constraint: Helices and perversions}
\author{D. Riccobelli$^{1,\,\star}$}
\email{davide.riccobelli@polimi.it}
\author{G. Noselli$^1$}
\email{giovanni.noselli@sissa.it}
\author{A. DeSimone$^{1,\,2}$}
\email{desimone@sissa.it}
\address[1]{SISSA -- International School for Advanced Studies, 34136 Trieste, Italy.}
\address[2]{The BioRobotics Institute and Department of Excellence in Robotics and A.I., Sant’Anna School of Advanced Studies, 56127 Pisa, Italy.}
\address[$\star$]{Current address: MOX -- Dipartimento di Matematica, Politecnico di Milano, 20133 Milano, Italy.}

\begin{abstract}
Mechanical instabilities can be exploited to design innovative structures, able to change their shape in the presence of external stimuli. In this work, we derive a mathematical model of an elastic beam subjected to an axial force and constrained to smoothly slide along a rigid support, where the distance between the rod midline and the constraint is fixed and finite.
Using both theoretical and computational techniques, we characterize the bifurcations of such a mechanical system, in which the axial force and the natural curvature of the beam are used as control parameters. We show that, in the presence of a straight support, the rod can deform into shapes exhibiting helices and perversions, namely transition zones connecting together two helices with opposite chirality. The mathematical predictions of the proposed model are also compared with some experiments, showing a good quantitative agreement. In particular, we find that the buckled configurations may exhibit multiple perversions and we propose a possible explanation for this phenomenon based on the energy landscape of the mechanical system.
\end{abstract}

\maketitle

\section{Introduction}

Bifurcation theory plays a crucial role in both theoretical and applied mechanics. One of the simplest examples is the buckling of a slender beam subject to an axial load: as the load reaches a critical value, the straight configuration loses stability and two other solutions appear. 
The bifurcation threshold is given by
\begin{equation}
\label{eq:Euler}
F_\text{cr}=\frac{B \pi^2}{k L^2}
\end{equation}


\maketitle
\noindent 
where $B$ is the lowest bending stiffness of the rod, $L$ its length and $k$ a numerical factor depending on the boundary conditions, also called \emph{column effective length factor}. This formula, first obtained by L.~Euler in 1757, is of fundamental importance in applied engineering in order to predict the collapse of slender structures.

Historically, the buckling of solid bodies has been considered exclusively in view of its connection with the failure of slender structures under compression, hence as something to avoid. Recently, a new paradigm has emerged \cite{Reis_2015} such that the buckling of elastic structures is now exploited as a possible mechanism to control the shape of elastic solids, triggering instabilities on demand through the application of external stimuli. In this context, the control of shape through buckling has important applications in robotics \cite{Wang_2018,Yeh_2019,Su_2020} and in the design of innovative materials \cite{Mullin_2007,bertoldi2010negative,florijn2014programmable,Caruso_2018}.

Among the different morphologies that a filamentary structure can assume, \emph{helices} are possibly the simplest and most interesting. In fact, many biological structures exhibit helical shapes, such as the DNA molecules \cite{watson_molecular_1953,Bouchiat_1998,Thompson_2002}, climbing plants and tendrils \cite{Goriely_2006}, bacteria flagella \cite{Chattopadhyay_2006,Rodenborn_2013}, and human hairs \cite{Bertails_2006,Miller_2014}. Furthermore, helical structures display peculiar mechanical properties \cite{Chouaieb_2006} which are exploited in advanced engineering applications such as the design of artificial muscles \cite{Haines_2016}, metamaterials \cite{noselli2019smart,riccobelli2020mechanics} and deployable structures \cite{Pellegrino_2001,olson2013deployable,Lachenal_2012,Lessinnes_2016}.\\
It is not rare to observe filamentary structures exhibiting helices of opposite chirality connected by a transition zone called \emph{perversion} \cite{Goriely_1998}. Some examples of objects that frequently exhibit perversions are telephone chords and plant tendrils. Mathematically, a perversion in an isolated rod can be described as a heteroclinic solution of the equilibrium equations, connecting two fixed points corresponding to the asymptotic helices \cite{mcmillen2002tendril}. In recent years, a significant effort has been devoted to understand the mathematics and mechanics of perversions, even though some peculiarities of these structures are still unclear, such as the appearance of multiple perversions (see for instance \cite{Domokos_2005,Lestringant_2017,Wang_2020}).

In this paper, we investigate the bifurcations of a simple mechanical structure composed of a beam that can smoothly slide along a rigid support. Differently from  other studies \cite{Cicconofri_2015,Dal_Corso_2017}, we consider the case in which the rod midline is at a finite distance from the support. This peculiar geometry is suggested by recent studies of assemblies of interlocked strips, able to slide along their common edge, inspired by unicellular protists (Euglenids, see \cite{arroyo2014shape,noselli2019smart,riccobelli2020mechanics}).
Since a full analysis of these structures is not available at present, we consider here a simplified case in which the edge of a flexible rod slides along a rigid constraint. While the proposed model contributes to the understanding of structural systems inspired by the euglenoid pellicle, it also provides innovative solutions for the design of tunable compliant mechanisms or propulsive, helical appendages in robotic applications.

The work is organized as follows: in section~\ref{sec:model} we construct the mathematical model as based on the Cosserat theory of rods. In section~\ref{sec:straight}, we specialize the model to the case of a rectilinear support, and find that the equilibrium equations admit non trivial explicit solutions. In section~\ref{sec:stability}, we use perturbation techniques to study the stability of the straight configuration and characterize the behavior of the bifurcation near the stability threshold. In section~\ref{sec:numerics}, we perform a numerical approximation of the fully nonlinear equations to study the post-buckling evolution of the equilibrium configuration. We also present some experimental results to validate the predictions of the mathematical model. Finally, we summarize in section~\ref{sec:conclusions} the main results together with some concluding remarks.

\section{Rods hooked to a smooth rigid constraint}
\label{sec:model}
In this section, we characterize the kinematics of an elastic rod hooked to a rigid curve. In particular, the distance of each point of the rod midline from the support is fixed but the beam can slide without friction along the curve.

\subsection{Kinematics}
We model the beam as a special Cosserat rod \cite{LAntman} embedded in the three dimensional Euclidean space $\mathbb{E}^3$. Let
\[
\vect{r}_0(s) = s \vect{E}_3,\qquad \vect{d}^0_i = \vect{E}_i,\qquad i=1,\,2,\,3,
\]
be the reference configuration of the rod, where $(\vect{E}_1,\,\vect{E}_2,\,\vect{E}_3)$ is the canonical vector basis in the reference frame and $\vect{d}_i^0$ are the directors of the beam. We denote by
\[
\vect{r},\,\vect{d}_i:[0,\,L]\rightarrow\R^3,\qquad i=1,\,2,\,3,
\]
the actual configuration of the rod, such that $\vect{r}$ is the function representing its midline and $\vect{d}_i$ are the three orthonormal directors. We assume that the beam is unshearable and inextensible, leading to the constraint
\begin{equation}
\label{eq:inex}
\vect{r}' = \vect{d}_3,
\end{equation}
where a prime denotes the derivative with respect to $s$.  Since $|\vect{d}_3|=1$, from \eqref{eq:inex}  we have that $s$ is the arclength also of the deformed midline $\vect{r}$. The unshearability of the rod is guaranteed by the fact that $\vect{r}'$ and $\vect{d}_3$ share the same direction.

\begin{figure}[t]
\centering
\includegraphics[height=0.3\textwidth]{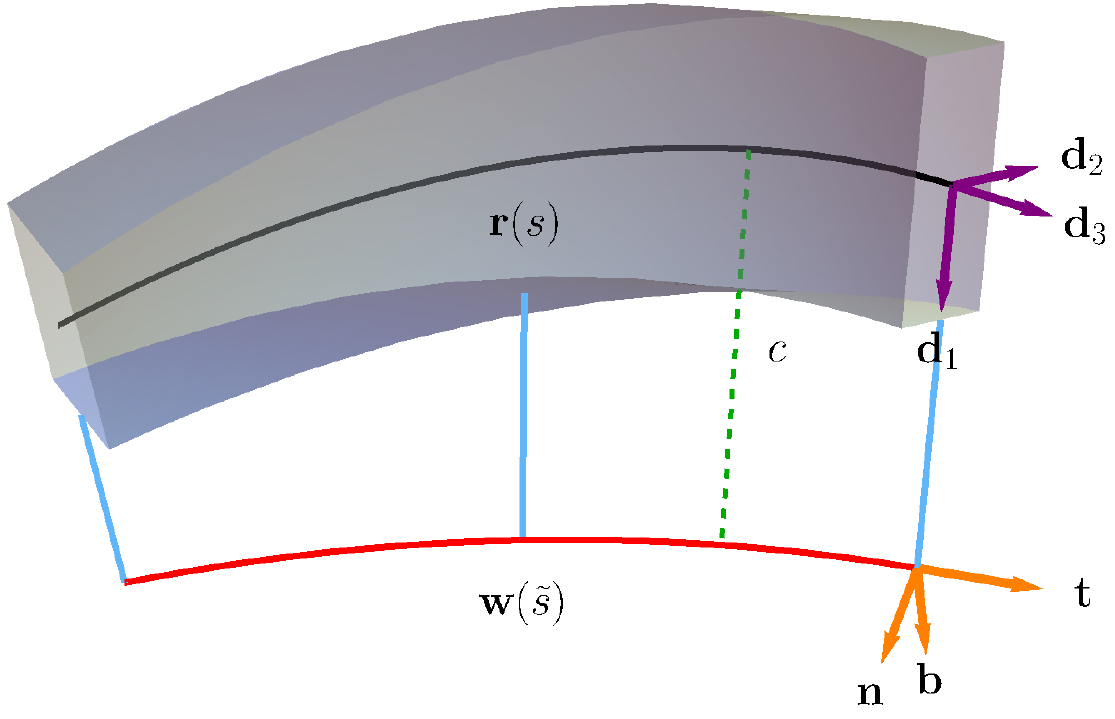}\quad
\includegraphics[height=0.3\textwidth]{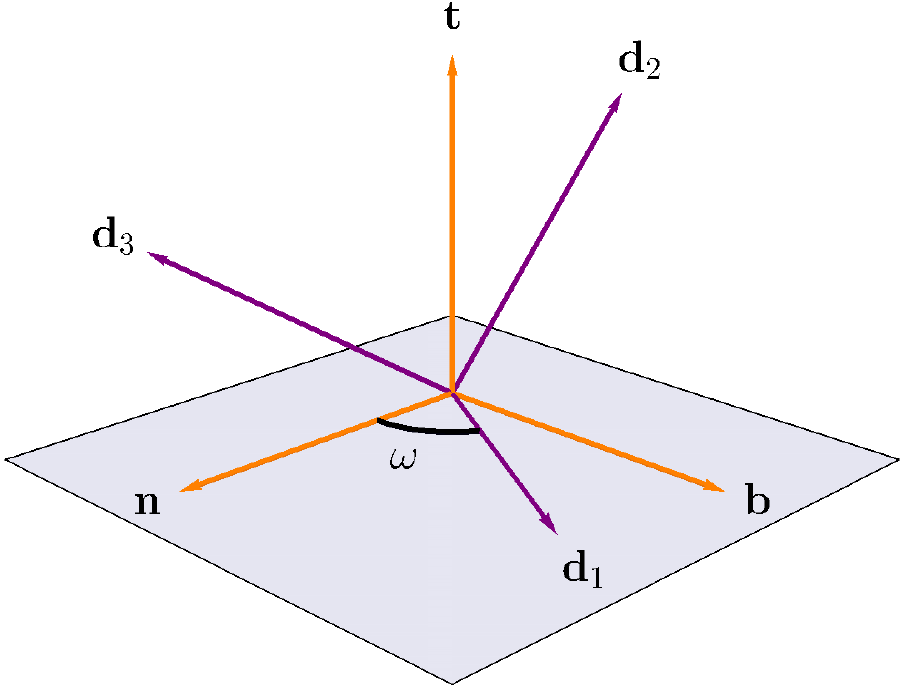}
\caption{(Left) Representation of the rod and of its support, together with the directors and the Serret-Frenet basis. (Right) Representation of the angle $\omega$; we remark that $\vect{n},\,\vect{b}$ and $\vect{d}_1$ are all contained in the same plane showed in the picture.}
\label{fig:conf}
\end{figure}

The constraint between the rod and the support, described by the curve $\vect{w}:\R\rightarrow\R^3$, is realized through a series of rigid connectors.
For the sake of simplicity, we assume that the connectors are continuously distributed along the beam. They are directed from the rod midline according to the vector field $\vect{c}(s)$ and can freely slide along the constraint $\vect{w}$, identifying a curve $\tilde{\vect{r}}(s) = \vect{r}(s)+\vect{c}(s)$.
We now enforce a compatibility constraint, analogous to the one exploited in \cite{noselli2019smart,riccobelli2020mechanics}: this is such that the image of $[0,\,L]$ through $\tilde{\vect{r}}$ must be contained in the image of $\vect{w}$, more explicitly
\begin{equation}
\label{eq:compatibility_single}
\vect{w}(\tilde{s}) = \vect{r}(s)+\vect{c}(s),
\end{equation}
for some $\tilde{s}=f(s)$, where $\tilde{s}$ is the arclength of the curve $\vect{w}$.
Since the connectors cannot self intersect, we enforce that $f$ is at least continuous and invertible.
In the following, we assume that the norm of $\vect{c}$ is constant with respect to $s$ and that $\vect{c}$ is aligned with the director $\vect{d}_1$, namely,
$
\vect{c}(s) = c\vect{d}_1(s).$
Then, the compatibility equation \eqref{eq:compatibility_single} modifies into 
\begin{equation}
\label{eq:compatibility_single_f}
\vect{w}(f(s)) = \vect{r}(s) + c\vect{d}_1(s), 
\end{equation}
see Figure~\ref{fig:conf} (left).
Assuming a sufficient regularity for all the functions, we can differentiate \eqref{eq:compatibility_single_f} to obtain
\begin{equation}
\label{eq:compatibility_differ}
\frac{d\vect{w}}{d\tilde{s}}(f(s))f'(s) = \vect{d}_3(s) +c\vect{d}_1'(s).
\end{equation}
Equation \eqref{eq:compatibility_differ} can be used to obtain some restrictions on the directors. First, we introduce the strain $\vect{u}$ \cite{LAntman}, defined as the vector function satisfying the following relation:
\begin{equation}
\label{eq:diu}
\vect{d}_i'(s) = \vect{u}(s)\times\vect{d}_i(s).
\end{equation}
Using this definition and computing the scalar product of \eqref{eq:compatibility_differ} with $\vect{d}_1$, we get
\begin{equation}
\label{eq:orth_d1}
f'(s)\frac{d\vect{w}}{d\tilde{s}}(f(s))\cdot\vect{d}_1(s) = 0,
\end{equation}
so that $\vect{d}_1$ must belong to the plane orthogonal to $d\vect{w}/d\tilde{s}$.
A vector basis for such a plane is provided by the orthonormal vectors $\vect{n}(\tilde{s})$ and $\vect{b}(\tilde{s})$, the normal and binormal unit vectors of the Serret--Frenet basis of the curve $\vect{w}(\tilde{s})$, defined as
\begin{equation}
\label{eq:Serret-Frenet}
\vect{t} = \frac{d\vect{w}}{d\tilde{s}},\qquad
\vect{n} = \left|\frac{d\vect{t}}{d\tilde{s}}\right|^{-1}\frac{d\vect{t}}{d\tilde{s}},\qquad
\vect{b} = \vect{t}\times\vect{n},
\end{equation}
whose derivatives satisfy the Serret-Frenet formulae
\begin{equation}
\label{eq:d_Serret-Frenet}
\frac{d\vect{t}}{d\tilde{s}} = \kappa\vect{n},\qquad
\frac{d\vect{n}}{d\tilde{s}} = -\kappa\vect{t}+\tau\vect{b},\qquad
\frac{d\vect{b}}{d\tilde{s}} = -\tau\vect{n},
\end{equation}
where $\kappa$ and $\tau$ are the curvature and the torsion of $\vect{w}$, respectively, and may depend on $\tilde{s}$.

Since $\vect{d}_1(s)$ is perpendicular to $\vect{t}(\tilde{s})$, see \eqref{eq:orth_d1}, there exist a function $\omega:[0,\,L]\rightarrow\R$ such that
\begin{equation}
\label{eq:d1omega}
\vect{d}_1(s) = \cos\omega(s)\vect{n}(\tilde{s}) + \sin\omega(s)\vect{b}(\tilde{s}),
\end{equation}
where we recall that $\tilde{s} = f(s)$ and $\omega(s)$ is the angle between $\vect{d}_1(s)$ and $\vect{n}(\tilde{s})$, see Figure~\ref{fig:conf} (right). Furthermore, the vector $\vect{d}_3(s)$ can be expressed in terms of $\omega$ and $f$ using the Serret-Frenet formulae \eqref{eq:Serret-Frenet} together with relation \eqref{eq:compatibility_differ}, obtaining
\begin{equation}
\label{eq:d3_f}
\vect{d}_3 = \frac{d\vect{w}}{d\tilde{s}}f'  - c\vect{d}_1'= \left(1+c\kappa \cos(\omega)\right)f'\vect{t}+c \sin(\omega)\left(\omega' + \tau f'\right)\vect{n}-c \cos(\omega)\left(\omega' + \tau f'\right)\vect{b}.
\end{equation}
Enforcing the inextensibility constraint \eqref{eq:inex}, we can use \eqref{eq:d3_f} to get a differential equation for $f$
\begin{equation}
\label{eq:inex_f}
|\vect{d}_3|^2=f'^2 \left(c^2 \tau ^2+(c \kappa  \cos (\omega )+1)^2\right)+2 c^2 \tau  f' \omega '+c^2 \omega '^2=1.
\end{equation} 
Finally, $\vect{d}_2$ can be expressed as a function of $\omega$ by computing the vector product of $\vect{d}_3$ and $\vect{d}_1$, obtaining
\begin{equation}
\label{eq:d2omega}
\vect{d}_2(s) =c \left(\tau  f'+\omega '\right)\vect{t}-f'\sin (\omega) (c \kappa \cos (\omega)+1)\vect{n} +f' \cos (\omega ) (c \kappa \cos (\omega)+1)\vect{b}.
\end{equation}

Summing up, we have used the compatibility constraint \eqref{eq:compatibility_single} to relate the directors of the rod with the Serret-Frenet frame. In particular, we have defined the angle $\omega$ as the angle identified by the vectors $\vect{n}$ and $\vect{d}_1$. Finally, we have expressed all the directors as functions of the kinematic variable $\omega$ (see \eqref{eq:d1omega}, \eqref{eq:d3_f} and \eqref{eq:d2omega}). In the following, we will use these results to derive the strain energy as a functional depending only on $\omega$.

\subsection{Strain measures and constitutive assumptions}

We can express the strain vector $\vect{u}$ as a function of $\omega$ by using \eqref{eq:diu} and the kinematics developed in the previous section. More explicitly, from \eqref{eq:compatibility_differ}, we obtain
\[
\vect{d}_1'=-u_2\vect{d}_3 + u_3\vect{d}_2 = c^{-1}(f'\vect{t} - \vect{d}_3),
\]
so that, computation of the dot product of the equation above with the directors $\vect{d}_2$ and $\vect{d}_3$, yields
\begin{equation}
\label{eq:u23}
u_2 = c^{-1}\left(1-f'\vect{t}\cdot\vect{d}_3\right)=\frac{1-f'^2 (c \kappa \cos (\omega)+1)}{c},\qquad
u_3 = c^{-1}f'\vect{t}\cdot\vect{d}_2=f' \left(\tau  f'+\omega '\right).
\end{equation}
We can use again \eqref{eq:diu} to express $u_1$ as a function of $\omega$. Since $u_1=-\vect{d}_3'\cdot\vect{d}_2$, we obtain:
\begin{equation}
\label{eq:u1}
\begin{multlined}
u_1 = c^2 \kappa ^3 f'^3 \sin (\omega ) \cos ^2(\omega )+\kappa \Bigg(f' \sin (\omega ) \left(3 c^2 \tau  f' \omega '+f'^2 \left(c^2 \tau ^2+1\right)+2 c^2 \omega '^2\right)+\\
+c^2 \cos (\omega ) \left(-f'' \omega '+f'^3 \frac{d\tau}{d\tilde{s}}+f' \omega ''\right)\Bigg)+c \kappa ^2 f'^3 \sin (2 \omega )+\\
+c \left(-c f'^2 \cos (\omega ) \frac{d\kappa}{d\tilde{s}} \left(\tau  f'+\omega '\right)-f'' \omega '+f'^3 \frac{d\tau}{d\tilde{s}}+f' \omega ''\right).
\end{multlined}
\end{equation}

Having fully characterized the kinematics of the sliding rod in terms of $\omega$, we proceed by introducing some constitutive assumption. We assume that the rod is hyperelastic and, in particular, we use the Kirchhoff's strain energy functional \cite{audoly2010elasticity}
\begin{equation}
\label{eq:energy_one_rod}
\mathcal{W} = \int_0^L W(\omega,\,\omega',\,\omega'')\,ds =\frac{1}{2}\int_0^L B_1(u_1 - u_1^{\star})^2+B_2(u_2 - u_2^{\star})^2+ T(u_3 - u_3^{\star})^2\,ds,
\end{equation}
where $B_1$ and $B_2$ are the bending stiffnesses with respect to the directions $\vect{d}_1$ and $\vect{d}_2$, respectively, and $T$ is the torsional stiffness. The quantities $-u_1^{\star}$, $-u_2^{\star}$ and $-u_3^{\star}$ represent the flexural and torsional strains of the reference configuration relative to the stress free configuration. We will refer to $u_j^{\star}$ as the \emph{natural} or \emph{intrinsic} curvatures and twist of the beam.

\section{Rod subject to an axial load and connected to a straight constraint}
\label{sec:straight}
We specialize the model developed in the previous section by considering a rectilinear support, such that $
\vect{w}(\tilde{s}) = \tilde{s}\vect{e}_3,$
where $(\vect{e}_1,\,\vect{e}_2,\,\vect{e}_3)$ is the canonical vector basis in the actual configuration.
In this case, the normal and the binormal vector to $\vect{w}$ are not uniquely defined. We set $\vect{n} = \vect{e}_1$, so that $\vect{b} = \vect{e}_2$ and both $\kappa$ and $\tau$ vanish.  In such a case, the differential equation \eqref{eq:inex_f} is much simpler and reduces to
\begin{equation}
\label{eq:fprimostraight}
f'(s) = \pm\sqrt{1-c^2 \omega '(s)^2}.
\end{equation} 
We observe that the invertibility of $f$ implies that the function must be strictly monotone and from \eqref{eq:fprimostraight} we have
\[
-\frac{1}{c}\leq\omega'\leq\frac{1}{c}.
\]

We assume that the rod cannot slide along $\vect{w}$ in $s=0$. We enforce this constraint by setting $f(0)=0$ as initial condition for the differential equation \eqref{eq:fprimostraight}. Without loss of generality, we assume that $f'(s)\geq0$ as due to the symmetries of the problem. Using \eqref{eq:fprimostraight}, we rewrite the components of the strain vector, given by \eqref{eq:u23}-\eqref{eq:u1}, into the following form
\begin{equation}
\label{eq:strain_u}
u_1 = \frac{c \omega ''}{\sqrt{1-c^2 \omega '^2}},\qquad u_2 =c \omega '^2,
\qquad u_3 = \omega ' \sqrt{1-c^2 \omega '^2}.
\end{equation}
We use \eqref{eq:strain_u} to write the strain energy density of the rod \eqref{eq:energy_one_rod} as a function of $\omega$ and its derivatives:
\begin{equation}
\label{eq:ener_omega}
\begin{multlined}
W=\frac{1}{2}\Bigg[B_1 \left(\frac{c \omega ''}{\sqrt{1-c^2 \omega '^2}}-u_1^{\star}\right)^2+B_2 \left(c \omega '^2-u_2^{\star}\right)^2+T \left(\omega ' \sqrt{1-c^2 \omega '^2}-u_3^{\star}\right)^2\Bigg].
\end{multlined}
\end{equation}

We assume that the beam is loaded at $\vect{r}(L)$ by a force $\vect{F}=F\vect{e}_3$. Exploiting the expression \eqref{eq:d3_f}, we compute the actual height of the deformed rod
\[
\vect{r}(L)\cdot\vect{e}_3 = \int_0^L\vect{d}_3\cdot\vect{e}_3\,ds = \int_0^L \sqrt{1-c^2\omega'^2}\,ds,
\]
so that the work done by the external forces reads
\[
\mathcal{P}[\omega] =\int_0^L P(\omega')\,ds=  F\int_0^L  \sqrt{1-c^2\omega'^2}\,ds - L.
\]
We remark that the rigid support provides a frictionless constraint to the rod, thus the work of the reaction forces exerted on the beam is zero. The absence of friction between the rod and the support is of course an idealization and the study of the effect of friction is beyond the scope of this article.

We are now in the position to define the total energy functional $\Psi$ as
\begin{equation}
\Psi[\omega] = \mathcal{W}[\omega]- \mathcal{P}[\omega],
\end{equation}
whose first variation reads
\begin{equation}
\label{eq:first_var}
\begin{multlined}
\delta\Psi(\omega)[\delta\omega] = \int_0^L \left[\frac{d^2}{ds^2}\frac{\partial W}{\partial\omega''}-\frac{d}{ds}\frac{\partial W}{\partial \omega'}+\frac{d}{ds}\frac{\partial P}{\partial \omega'}\right]\delta\omega\,ds+\\
+\left[\left(\frac{\partial W}{\partial \omega'}-\frac{\partial P}{\partial \omega'}-\frac{d}{ds}\frac{\partial W}{\partial\omega''}\right)\delta\omega\right]_{s=0}^{s=L}+\left[\frac{\partial W}{\partial\omega''}\delta\omega'\right]_{s=0}^{s=L}.
\end{multlined}
\end{equation}
We obtain the balance equation of the mechanical system, together with the natural boundary conditions, by requiring that the energy functional $\Psi$ be stationary. In particular, the first term of \eqref{eq:first_var} provides the Euler-Lagrange equation of the mechanical system, while the other two terms give the natural boundary conditions.
More explicitly, the Euler-Lagrange equation reads
\begin{equation}
\label{eq:EL_nonlinear}
\begin{multlined}
f'^2 \omega '' \left(4 B_{1} c^4 \omega''' \omega '+\left(c^2 \omega '^2-1\right)^2 \left(6 c^2 (T-B_{2}) \omega '^2+2 B_{2} c u_2^{\star}-T\right)\right)+\\
+B_{1} c^2 \omega'''' f'^4+B_{1} c^4 \omega ''^3 \left(3 c^2 \omega '^2+1\right)+c^2 f'^3 \omega '' \left(T u_3^{\star} \omega ' \left(2 c^2 \omega '^2-3\right)-F\right)=0,
\end{multlined}
\end{equation}
where $f'$ is given by \eqref{eq:fprimostraight}, while
the boundary terms are obtained as
\begin{equation}
\label{eq:natural_BC}
\left\{
\begin{aligned}
&\begin{aligned}
\frac{\partial W}{\partial \omega'}-\frac{\partial P}{\partial \omega'}-\frac{d}{ds}\frac{\partial W}{\partial\omega''}=&-\frac{B_{1} c^4 \omega ' \omega ''^2}{f'^4}-\frac{B_{1} c^2 \omega'''}{f'^2}+c \omega ' \left(c (2 B_{2}-T) \omega '^2-2 B_{2} u_2^{\star}\right)+\\
&+\frac{c^2 \omega ' \left(F+T u_3^{\star} \omega '\right)}{f'}-T u_3^{\star} f'+T f'^2 \omega ',
\end{aligned}\\
&\frac{\partial W}{\partial\omega''}=\frac{B_{1} c \left(c \omega ''-u_1^{\star} f'\right)}{f'^2}.\\
\end{aligned}
\right.
\end{equation}

In the following, we discuss the behavior of the system subject to different boundary conditions. 

\subsection{Boundary conditions}

We distinguish two cases:
\begin{itemize}
\item \textbf{Case A -- free ends}. We assume that the rod is free to rotate about the support at both ends. We remark that both the Euler-Lagrange equation \eqref{eq:EL_nonlinear} and the natural boundary conditions \eqref{eq:natural_BC} depend on $\omega$ only through its derivatives. This is due to the fact that the rod can rigidly rotate about the support without storing mechanical energy. To rule out these rigid motions, we set $\omega(0)=0$. The problem is complemented by the natural boundary conditions
\begin{equation}
\label{eq:BC_caso_A}
\left\{
\begin{aligned}
&\left.\left(\frac{\partial W}{\partial \omega'}-\frac{\partial P}{\partial \omega'}-\frac{d}{ds}\frac{\partial W}{\partial\omega''}\right)\right|_{s=L}=0,\\
&\left.\frac{\partial W}{\partial\omega''}\right|_{s=0,\,L}=0,
\end{aligned}
\right.
\end{equation}
which arise from the first variation of the energy, see \eqref{eq:first_var}-\eqref{eq:natural_BC}.
\item \textbf{Case B -- pinned ends}. We assume that $\omega(0)=\omega(L)=0$, while we leave the first derivative of $\omega$ unconstrained. In analogy with the Euler's buckling problem, we refer to these boundary conditions as the pinned end case.
As for the natural boundary conditions, from \eqref{eq:first_var} we get:
\begin{equation}
\label{eq:BC_caso_B}
\left.\frac{\partial W}{\partial\omega''}\right|_{s=0,\,L}=0.
\end{equation}
\end{itemize}
The analysis of Case A is reported in detail in the following sections, and can be easily adapted to Case B. For the sake of brevity, we will report the results of the stability analysis for the pinned end case in a separate section, omitting the explicit calculations and highlighting the few differences resulting from the different choice of boundary conditions.

\subsection{Non-dimensionalization of the equilibrium equations}

Before proceeding with the analysis of the model, we non-dimensionalize the system with respect to $c$, namely the distance of the rod midline from the support, and the bending stiffness $B_2$. The dimensionless counterpart of the physical quantities introduced so far are
\begin{equation}
\label{eq:nondim}
\begin{gathered}
\hat{s} = \frac{s}{c}\qquad \hat{L}= \frac{L}{c}\qquad \beta = \frac{B_1}{B_2}\qquad \sigma = \frac{T}{B_2} \qquad \hat{u}_j^{\star} = u_j^{\star} c,\qquad\hat{F} =\frac{c^2 F}{B_2} \\
\hat{\Psi} = \frac{c \Psi}{B_2}\qquad\hat{\mathcal{W}} = \frac{c \mathcal{W}}{B_2}\qquad\hat{\mathcal{P}} = \frac{c \mathcal{P}}{B_2}\qquad \hat{W} = \frac{c^2 W}{B_2}\qquad \hat{P} = \frac{c^2 P}{B_2}
\end{gathered}
\end{equation}

In the following, we will consider rods having a rectangular cross-section.
In particular, denoting by $h$ the width of the rod along $\vect{d}_1$ and by $t$ the thickness along $\vect{d}_2$, for $h> t$ it is a classical result that
\[
B_1 = \frac{Eht^3}{12},\qquad B_2 = \frac{Eh^3t}{12},\qquad T = \frac{G\chi ht^3}{3},
\]
where $E$ and $G$ are the Young's and shear moduli, respectively, and $\chi$ is a numerical factor depending on the aspect ratio $h/t$, tending to $1$ as $h/t$ tends to infinity. From the equation above, we get
\begin{equation}
\label{eq:betasigmarect}
\beta = \frac{t^2}{h^2},\qquad \sigma = \frac{t^2}{h^2}\frac{4 G \chi}{E} = \frac{t^2}{h^2}\frac{2\chi}{1 + \nu},
\end{equation}
where $\nu$ is the Poisson's ratio. We remark that for $h/t\gg 1$ it is reasonable to assume $\chi = 1$.

From now on, we will use only the dimensionless counterpart of the physical quantities, unless explicitly specified. For convenience, we drop the circumflex of the non-dimensional quantities defined in \eqref{eq:nondim}. In the next section, we discuss some analytical solutions of the Euler-Lagrange equation.

\subsection{Analytical solutions}
\label{sec:analytical}

\begin{figure}[t!]
\centering
\includegraphics[width=0.3\textwidth]{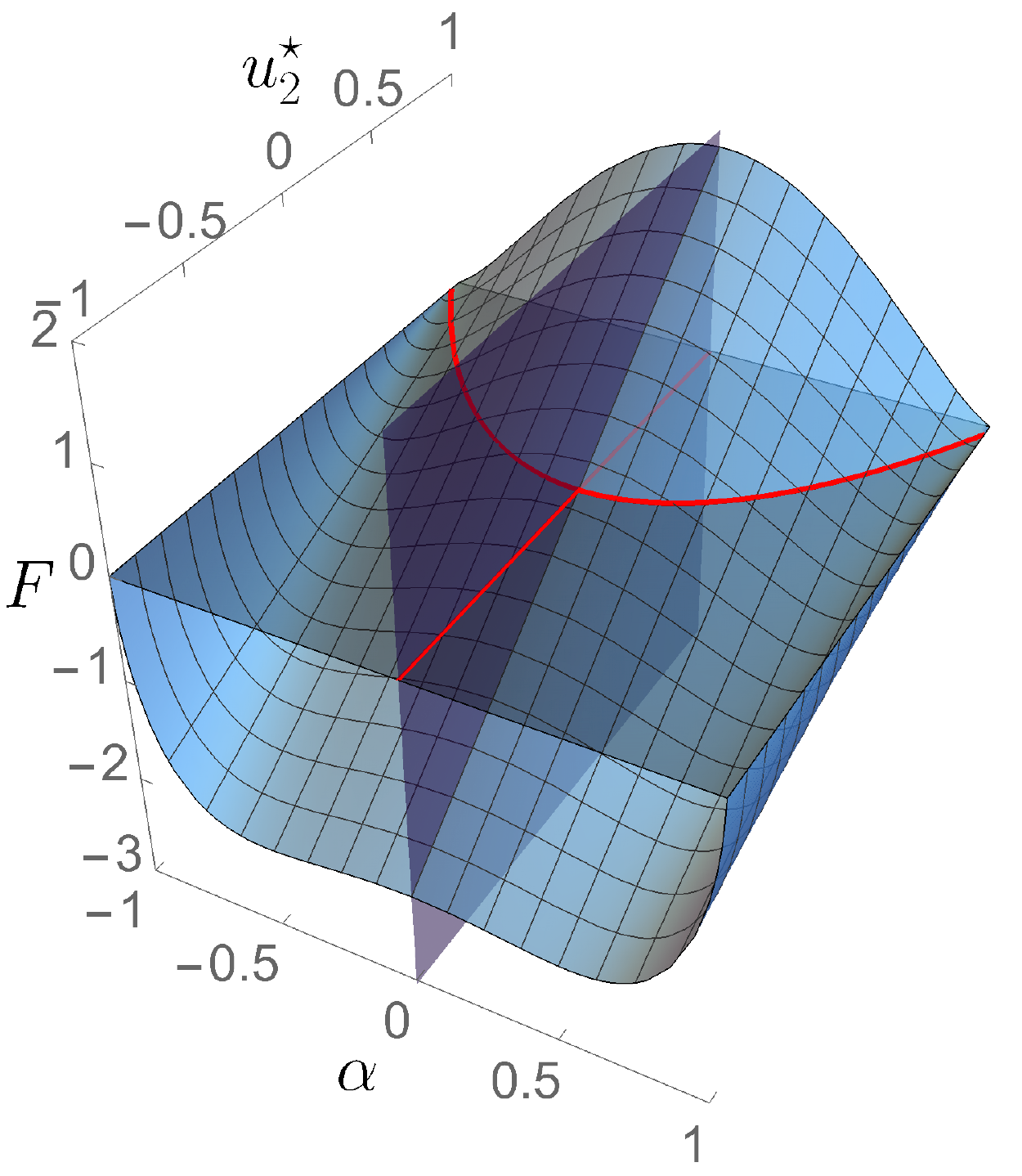}\hfill
\includegraphics[width=0.3\textwidth]{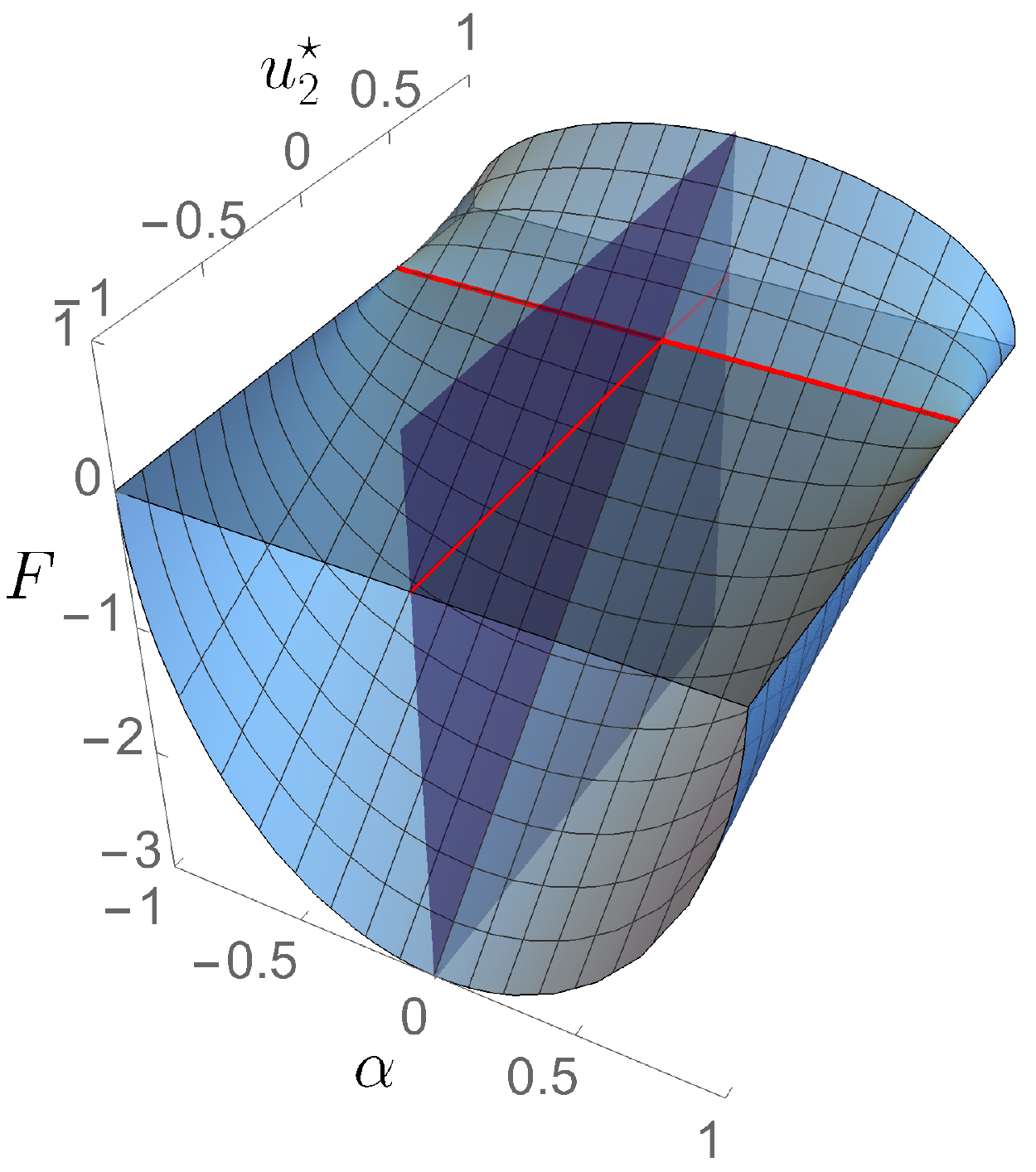}\hfill
\includegraphics[width=0.3\textwidth]{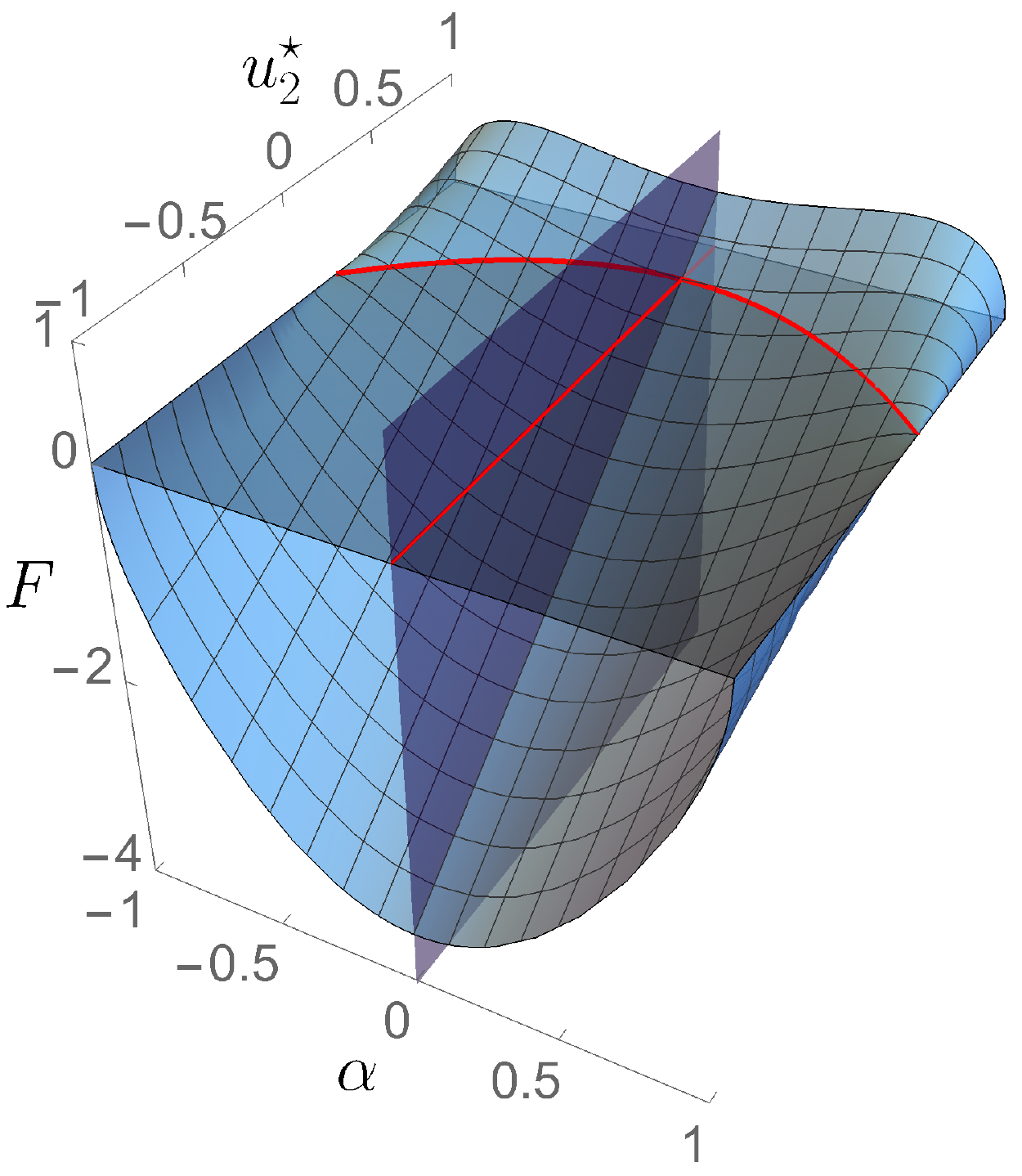}
\caption{Plot of the equilibrium force $F$ for helical configurations when $u_3^{\star}=0$ and $\sigma=0.1$ (left), $\sigma=1$ (center) and $\sigma=1.5$ (right). The purple plane $\alpha=0$ corresponds to the reference configuration, which is in equilibrium for all values of $F$ and $u_2^{\star}$. The red curves represent the bifurcation diagram obtained for $F=0$, where $u_2^{\star}$ plays the role of the control parameter. We remark that, if $\sigma=1$ and $u_2^{\star}=0.5$, all the helical configuration are in mechanical equilibrium for $F=0$ (center).}
\label{fig:3D}
\end{figure}

\begin{figure}[t!]
\centering
\includegraphics[width=0.33\textwidth]{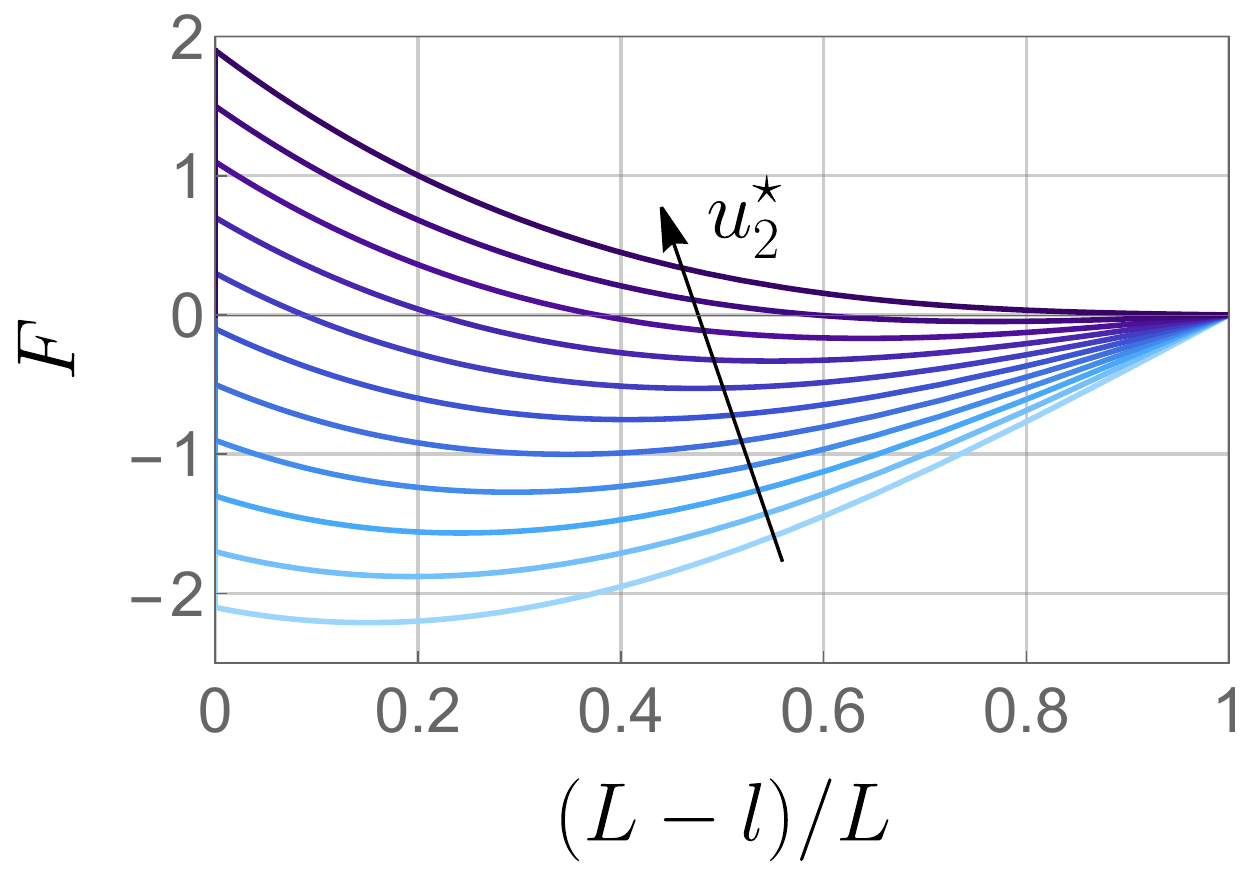}%
\includegraphics[width=0.33\textwidth]{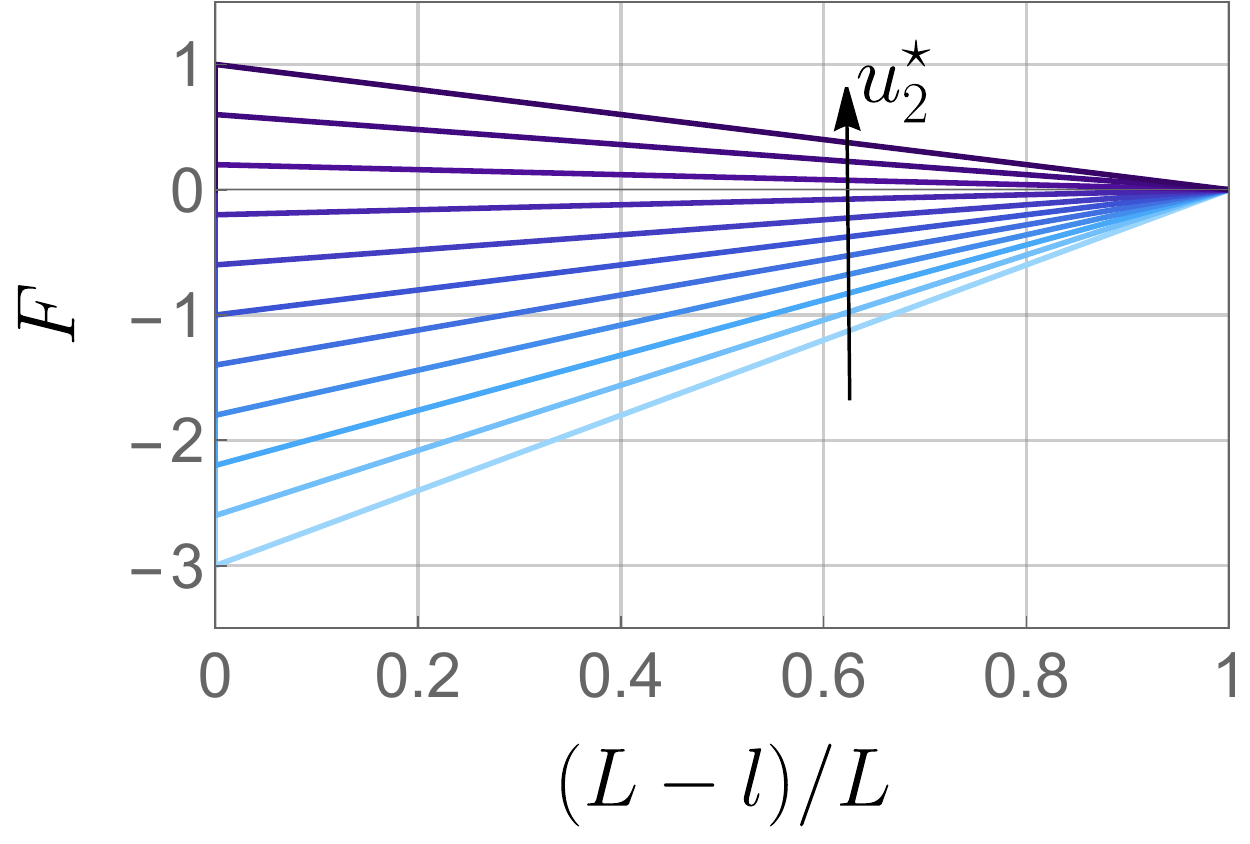}%
\includegraphics[width=0.33\textwidth]{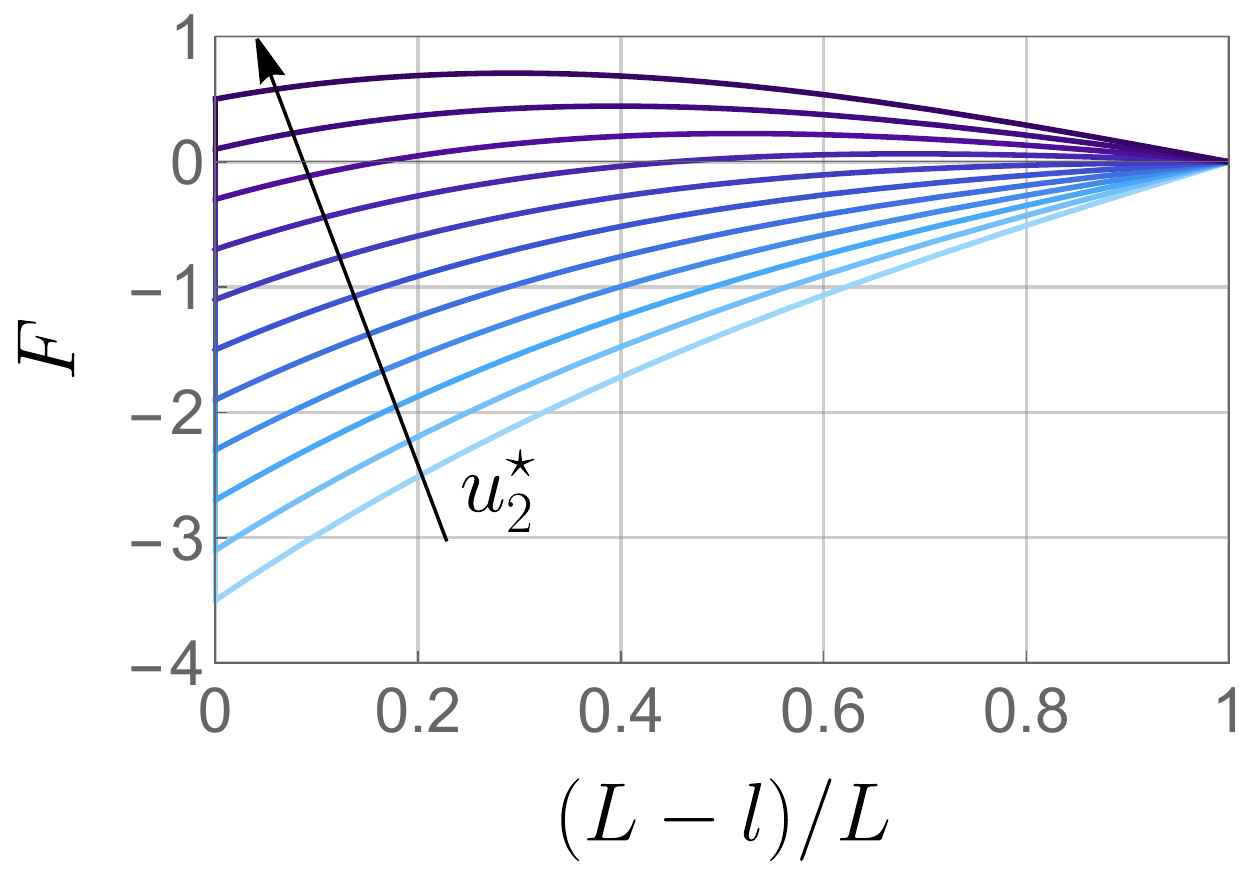}
\caption{Applied force $F$ versus the overall strain $(L-l)/L$ for $\sigma = 0.1$ (left), $\sigma=1$ (center), $\sigma=1.5$ (right). In the plots, we show several curves for $u_2^{\star}$ ranging from $-1$ to $1$ by steps of $0.2$. The arrows denote the direction in which $u_2^{\star}$ increases.}
\label{fig:force_disp}
\end{figure}

We set $\omega(0)=0$ together with the natural boundary conditions \eqref{eq:BC_caso_A}. As a result, the reference, straight configuration given by $\omega = 0$ is a solution whenever
\[
u_1^{\star} =u_3^{\star} = 0.
\]
We observe that this is not the only equilibrium configuration of the system. In fact, also 
\[
\omega(s) = \alpha s
\]
satisfies the Euler-Lagrange equation \eqref{eq:EL_nonlinear}. This solution corresponds to a helical configuration of the rod. The second natural boundary condition \eqref{eq:natural_BC} is automatically satisfied whenever $u_1^{\star}=0$, while the first one provides a relation between between $\alpha$, $F$, $u_2^{\star}$, and $u_3^{\star}$:
\begin{equation}
\label{eq:Feliche}
\alpha F=2 \sqrt{1-\alpha ^2} \alpha ^3 \sigma -\sqrt{1-\alpha ^2} \alpha  \sigma -2 \sqrt{1-\alpha ^2} \alpha ^3+2 \sqrt{1-\alpha ^2} \alpha  u_2^{\star}-2 \alpha ^2 \sigma  u_3^{\star}+\sigma  u_3^{\star}.
\end{equation}
We observe that, if $u_3^{\star}$ is zero in \eqref{eq:Feliche}, then the equation is satisfied for all $F$ and $u_2^{\star}$ for $\alpha=0$, which corresponds again to the reference configuration.
Instead, from $\alpha\neq0$ we obtain from \eqref{eq:Feliche}
\begin{equation}
\label{eq:F_alpha}
F = \sqrt{1-\alpha ^2} \left(2 \alpha ^2 \sigma -2 \alpha ^2-\sigma +2 u_2^{\star}\right).
\end{equation}
We show in figure~\ref{fig:3D} the three-dimensional plots of the equilibrium force $F$ as a function of $\alpha$ and $u_2^\star$ for $\sigma=0.1,\,1,\,1.5$.
In particular, using \eqref{eq:F_alpha}, we can compute the critical force at which the straight configuration buckles into a helical shape. In the limit $\alpha\rightarrow 0$, we get
\begin{equation}
\label{eq:F_cr}
F = -\sigma +2 u_2^{\star}.
\end{equation}
It is interesting to observe that, contrary to Euler's formula \eqref{eq:Euler}, the buckling load (and in general the force necessary to generate a helix with a given $\alpha$) is independent of the rod length, even though the structure undergoes a global buckling. We argue that this is because the strain variables are constant along the beam. In figure~\ref{fig:force_disp}, we show the applied force as a function of the overall strain of the rod in the helical configuration, defined as
\[
\frac{L-l}{L}\qquad\text{where }l=\vect{r}(L)\cdot\vect{e}_3=f(L)=L\sqrt{1-\alpha^2}.
\]
Another interesting feature of the buckling load becomes apparent if we rewrite \eqref{eq:F_cr} in dimensional variables. Using \eqref{eq:nondim} we get
\begin{equation}
\label{eq:dimensional_buckling_load}
F = -\frac{T}{c^2} + \frac{2 B_2 u_2^\star}{c},
\end{equation}
which is identical to the buckling load reported in \cite{noselli2019smart}, equation (29). In that work, the authors considered a cylindrical assembly of interlocked rods, able to deform into helical configurations. In their formula the connector distance $c$ is replaced by the reference radius of the cylinder. As we will discuss later, we argue that this similarity arises since, in the linearized setting, the radius of the assembly studied in \cite{noselli2019smart} does not change. Despite these similarities, we remark that the nonlinear equations for the two systems are not the same so that the post-buckling evolution is different.

\begin{figure}[t!]
\centering
\includegraphics[width=0.5\textwidth]{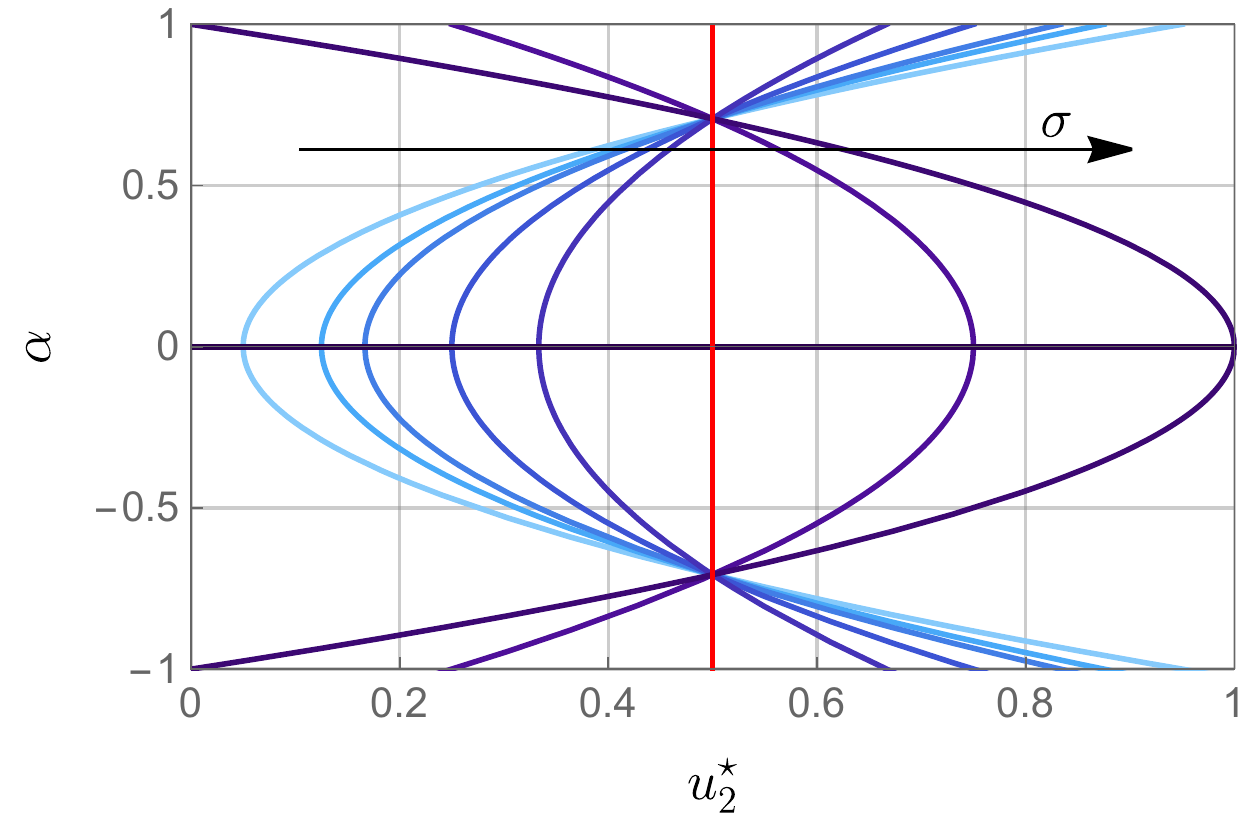}
\caption{Plots of $\alpha$ versus $u_2^{\star}$ for $\sigma=1/10,\,1/4,\,1/3,\,1/2,\,2/3,\,1 $ (red line)$,\,3/2,\,2$. The arrow denote the direction in which $\sigma$ increases.}
\label{fig:C}
\end{figure}

Finally, the base reference configuration undergoes a bifurcation for a critical value of $u_2^{\star}$ even in the absence of an external load, as one can notice from \eqref{eq:F_cr}. In such a case, the critical natural curvature is given by
\begin{equation}
\label{eq:u2scr}
u_2^{\star}=\frac{\sigma}{2}.
\end{equation}
In particular, \eqref{eq:F_alpha} admits an explicit expression for $\alpha$ in the absence of external forces
\[
\alpha=\pm\frac{\sqrt{\sigma -2 u_2^{\star}}}{\sqrt{2} \sqrt{\sigma -1}}.
\]
We plot the resulting bifurcation diagrams in figures~\ref{fig:3D}-\ref{fig:C}. We observe that the bifurcation is a supercritical or subcritical pitchfork depending on the sign of $\sigma - 1$. In the particular case in which $\sigma = 1$, all the helices are equilibrium solutions for $u_2^{\star}=1/2$ (see figure~\ref{fig:3D} center and figure~\ref{fig:C}).

\begin{remark*}
Summing up, if the rod does not have any natural twist (i.e. $u_3^\star=0$) or natural curvature about $\vect{d}_1$ (i.e. $u_1^\star=0$) but an arbitrary $u_2^\star$, then the straight reference configuration is in mechanical equilibrium. When the applied force is that given by \eqref{eq:F_alpha} a bifurcation of the equilibrium occurs, such that, in the post-buckling regime, the beam wraps about the support assuming a helical shape. In the absence of an external axial force, a bifurcation occurs if the natural curvature $u_2^{\star}$ is sufficiently high, according to relation~\eqref{eq:u2scr}.
\end{remark*}

In the next section, we perform a stability analysis, looking for other bifurcation points of the mechanical system.

\section{Stability analysis of the straight configuration}
\label{sec:stability}

In this section, we further investigate the stability of the straight, reference configuration using perturbation methods. First, we perform a linear stability analysis. Second, we classify the bifurcation points through a weakly nonlinear expansion, characterizing the initial evolution of the bifurcated solution in a neighborhood of the stability threshold.
Before proceeding with the linear analysis, let us introduce the following inner product between scalar functions, which is a rescaled version of the standard $L^2$ scalar product:
\begin{equation}
\label{eq:sp}
\langle f,\,g\rangle = \frac{2}{L}\int_0^Lfg\,ds.
\end{equation}

We consider two situations in which buckling is either triggered by the natural curvature $u_2^{\star}$ or by the axial force $F$. We denote by $\lambda$ the control parameter, namely either $u_2^{\star}$ or $-F$. 

\subsection{Linear analysis}
\label{sec:linear_analysis}
Let $\omega_b(s)$ be the bifurcated solution originating from the base solution $\omega_0(s)=0$ at $\lambda=\lambda_0$, so that
\[
\lim_{\lambda\rightarrow\lambda_0}\omega_b(s;\,\lambda)=0.
\]
We define the bifurcation mode $\omega_1$ as
\begin{equation}
\label{eq:buck_mode_def}
\omega_1 = \lim_{\lambda\rightarrow\lambda_0} \frac{\omega_b}{\|\omega_b\|},
\end{equation}
where $\|\cdot\|$ is the norm induced by the scalar product \eqref{eq:sp}, so that $\|\omega_1\|=1$. 
The bifurcated solution can be expressed in the following form
\[
\omega_b(s;\,\lambda) = \xi\omega_1(s)+v(s;\,\lambda),
\]
where $\xi$ is defined as $\xi = \langle \omega_b,\,\omega_1\rangle$, implying that
\begin{equation}
\label{eq:orthogonality}
\langle\omega_1,\,v\rangle = 0,
\end{equation}
and that $v = o(\xi)$ as $\lambda\rightarrow\lambda_0$.
Expanding \eqref{eq:EL_nonlinear} up to the first order in $\xi$, the linearized equation reads
\begin{equation}
\label{eq:ELlinear}
\beta  \omega_1''''(s)-(F+\sigma -2 u_2^{\star}) \omega_1''(s)=0,
\end{equation}
while the linearized form of the boundary terms \eqref{eq:natural_BC} are given by
\begin{equation}
\label{eq:naturalBClinear}
\left\{
\begin{aligned}
&\frac{\partial W}{\partial \omega'}-\frac{\partial P}{\partial \omega'}-\frac{d}{ds}\frac{\partial W}{\partial\omega''}=-\sigma  u_3^{\star}+\xi  \left((F+\sigma -2 u_2^{\star}) \omega_1'(s)-\beta  \omega_1'''(s)\right)+o(\xi),\\
&\frac{\partial W}{\partial\omega''}=-\beta u_1^\star+\beta  \xi  \omega_1''(s)+o(\xi).\\
\end{aligned}
\right.
\end{equation}
The general solution of the linear equation~\eqref{eq:ELlinear} is given by
\[
\omega_1 =c_1 + c_2 s +c_3 \sin\left(\frac{s \sqrt{2 u_2^{\star}-F-\sigma }}{\sqrt{\beta }}\right)+c_4 \cos\left(\frac{s \sqrt{2 u_2^{\star}-F-\sigma }}{\sqrt{\beta }}\right),
\]
if $2 u_2^{\star}-F \neq \sigma$, otherwise it is given by
\[
\omega_1 =c_1 + c_2 s +c_3 s^2+ c_4 s^3.
\]
As before, we first treat extensively the free ends case. We set both $u_1^{\star}$ and  $u_3^{\star}$ equal to zero, so that the straight reference configuration is in mechanical equilibrium. If $2 u_2^* - F = \sigma$, the second equation of \eqref{eq:BC_caso_A} together with \eqref{eq:naturalBClinear} provides
\[
c_3 = c_4 = 0.
\]
Since $\omega(0)=0$, $c_1=0$ and, since from \eqref{eq:buck_mode_def} $\|\omega_1\|=1$, $c_2=1$. In particular, we have obtained the helical configuration analyzed in the previous section.

If instead $2 u_2^* - F \neq \sigma$, using the first equation of \eqref{eq:naturalBClinear} and neglecting the remainder in \eqref{eq:BC_caso_A} we get $c_2=0$. Analogously, from the second equation of \eqref{eq:BC_caso_A} we get $c_4 = 0$ and
\begin{equation}
\label{eq:sol_pin_end}
2u_2^{\star}-F=\frac{\pi ^2 \beta  n^2}{L^2}+\sigma,
\qquad n\in\mathbb{N}\setminus\{0\},
\end{equation}
so that
\begin{equation}
\label{eq:omega_1_perversion}
\omega_1(s)=\sin \left(\frac{n \pi s}{L}\right),\qquad n\in\mathbb{N}\setminus\{0\},
\end{equation}
where $c_1$ vanishes since $\omega(0)=0$ and $c_3$ has been set equal to $1$ so that $\|\omega_1\|=1$. We observe that $\omega_1'$ assumes both positive and negative values in $(0,\,L)$, indicating that the rod coils about the support with opposite handedness, depending on the sign of $\omega'_1$. Following \cite{Goriely_1998}, we call perversion a point where $\omega'$ changes sign (i.e. the chirality of the rod changes).

We define the critical mode as the one corresponding to the lowest value of $2u_2^{\star}-F$. For the sinusoidal mode \eqref{eq:omega_1_perversion}, exhibiting $n$ perversions, the value of $2u_2^{\star}-F$ \eqref{eq:sol_pin_end} at which a bifurcation occurs is always higher than the one of the helical mode, see \eqref{eq:F_cr}.
As done in section \ref{sec:analytical}, if we rewrite the buckling load \eqref{eq:sol_pin_end} in dimensional variables, we get
\begin{equation}
\label{eq:dimensional_buckling_load2}
F = -\frac{T}{c^2} +\frac{2 B_2 u_2^\star}{c} - \frac{ B_1 n^2\pi^2}{L^2}.
\end{equation}
Once again, this buckling load is identical to the one reported in \cite{riccobelli2020mechanics}, section 5.1.1, for an axisymmetric assembly of interlocking and slidable beams. As mentioned for helices, this analogy follows from the fact that the distance of the rods from the symmetry axis does not change for small displacements, see (26) in \cite{riccobelli2020mechanics}. Thus, the  linearized kinematics of the two mechanical systems is very similar, where in particular the reference radius of the assembly plays the role of the connector distance $c$.

Interestingly, in the limit case of an infinite rod, the buckling load \eqref{eq:sol_pin_end} is independent of $n$ and coincides with the one obtained for the helices, see \eqref{eq:F_cr}. This might provide an explanation for the experimental observation of multiple perversions in similar mechanical systems \cite{mcmillen2002tendril,Domokos_2005,Lestringant_2017}, and we discuss this aspect into more detail in section \ref{sec:numerics_f}

The linear analysis proposed in this section allows us to predict the bifurcations of the mechanical system, without providing any information on the amplitude of the buckling mode.
In the following, we perform a weakly nonlinear stability analysis \cite{hutchinson1970postbuckling,Budiansky_1974} to classify the bifurcation points and characterize the post-critical configurations in a neighborhood of the bifurcation points.

\subsection{Weakly nonlinear analysis of the bifurcated branches exhibiting perversions}
\label{sec:WNL}
In this section, we focus our attention on the buckling modes exhibiting perversions, since we have already characterized the fully nonlinear behavior of the helical configurations in section~\ref{sec:analytical}.
In particular, we follow the approach proposed by Budiansky in \cite{Budiansky_1974}. 

We study the free ends case. Let $u_1^{\star} = u_3^{\star} = 0$, so that the reference configuration is in mechanical equilibrium.
Since $\omega_b$ is a solution of the problem \eqref{eq:EL_nonlinear}-\eqref{eq:BC_caso_B}, we get
\begin{equation}
\label{eq:stationarity_omega_b}
\delta\Psi(\omega_b;\,\lambda)[\delta\omega] = 0,
\end{equation}
where we have explicitly highlighted the dependence of the energy on the control parameter $\lambda$. We assume the following power series expansions
\begin{equation}
\label{eq:expansion_v_u_1^*}
\left\{
\begin{aligned}
&v = \xi^2\omega_2 + \xi^3\omega_3 + o(\xi^3),\\
&\lambda = \lambda_0 + \xi\lambda_1 + \xi^2 \lambda_2 + o(\xi^2).
\end{aligned}
\right.
\end{equation}

We remark that, since $\omega_1$ is a solution of the linearized problem \eqref{eq:ELlinear}, we have (see (4.7) in \cite{Budiansky_1974})
\begin{equation}
\label{eq:bif_criterion_budiansky}
\delta^2\Psi_c[\omega_1,\,\delta \omega]=0,
\end{equation}
for any admissible $\delta\omega$, where the subscript $c$ indicates that the second variation $\delta^2\Psi$ is computed in $(\omega_0 = 0,\,\lambda_0)$, i.e. 
\[
\delta^2\Psi_c[\omega_1,\,\delta \omega]\coloneqq\delta^2\Psi(0;\,\lambda_0)[\omega_1,\,\delta \omega].
\]
Using the ansatz \eqref{eq:expansion_v_u_1^*} and the following compact notation
\[
\delta^n\dot{\Psi}_c \coloneqq \left.\frac{d}{d\lambda}\delta^n\Psi(0,\,\lambda)\right|_{\lambda=\lambda_0}
\]
we can compute the expansion of \eqref{eq:stationarity_omega_b}, obtaining \cite{Budiansky_1974}
\begin{equation}
\label{eq:expansion_energy}
\begin{aligned}
\delta\Psi&(\omega_b;\,\lambda)[\delta\omega]=\bigg(\delta^2\Psi_c[\omega_2,\,\delta\omega]+\lambda_1\delta^2\dot{\Psi}_c[\omega_1,\,\delta\omega]+\frac{1}{2}\delta^3\Psi_c[\omega_1,\,\omega_1,\,\delta\omega]\bigg)\xi^2+\\
&+\bigg(\delta^2\Psi_c[\omega_3,\,\delta\omega]+\lambda_1\delta^2\dot{\Psi}_c[\omega_2,\,\delta\omega]+\lambda_2\delta^2\dot\Psi_c[\omega_1,\,\delta\omega]+\frac{1}{2}\lambda_1^2\delta^2\ddot{\Psi}_c[\omega_1,\,\delta\omega]+\\
&+\delta^3\Psi_c[\omega_1,\,\omega_2,\,\delta\omega]+\frac{1}{2}\lambda_1\delta^3\dot{\Psi}_c[\omega_1,\,\omega_1,\,\delta\omega]+\frac{1}{6}\delta^4\Psi_c[\omega_1,\,\omega_1,\,\omega_1,\,\delta\omega]\bigg)\xi^3+o(\xi^3)=0,
\end{aligned}
\end{equation}
where all the coefficients of the powers of $\xi$ must vanish to guarantee that the energy on the bifurcated branch be stationary.
\subsubsection{Weakly nonlinear analysis in the absence of an axial force}
\label{sec:WNL_u2s}
In this section, we assume that $F=0$ and we use $u_2^{\star}$ as control parameter of the bifurcation, i.e. $\lambda=u_2^\star$.
From \eqref{eq:sol_pin_end}, we have
\[
\lambda_0=\frac{\pi ^2 \beta  n^2}{2L^2}+\frac{\sigma}{2}.
\]
Setting the second order term of \eqref{eq:expansion_energy} equal to zero and $\delta\omega=\omega_1$, we get
\begin{equation}
\label{eq:lambda_1}
\lambda_1 = -\frac{1}{2}\frac{\delta^3\Psi_c[\omega_1,\,\omega_1,\,\omega_1]}{\delta^2\dot{\Psi}_c[\omega_1,\,\omega_1]}=0.
\end{equation}
This result follows from the fact that for the energy functional \eqref{eq:energy_one_rod} $\delta^3\Psi_c[v_1,\,v_2,\,v_3]=0$
for any $v_1,\,v_2,\,v_3$ and
\[
\delta^2\Psi_c[\omega_2,\,\omega_1] = \delta^2\Psi_c[\omega_1,\,\omega_2]=0.
\]

\begin{figure}[t!]
\includegraphics[width=0.33\textwidth]{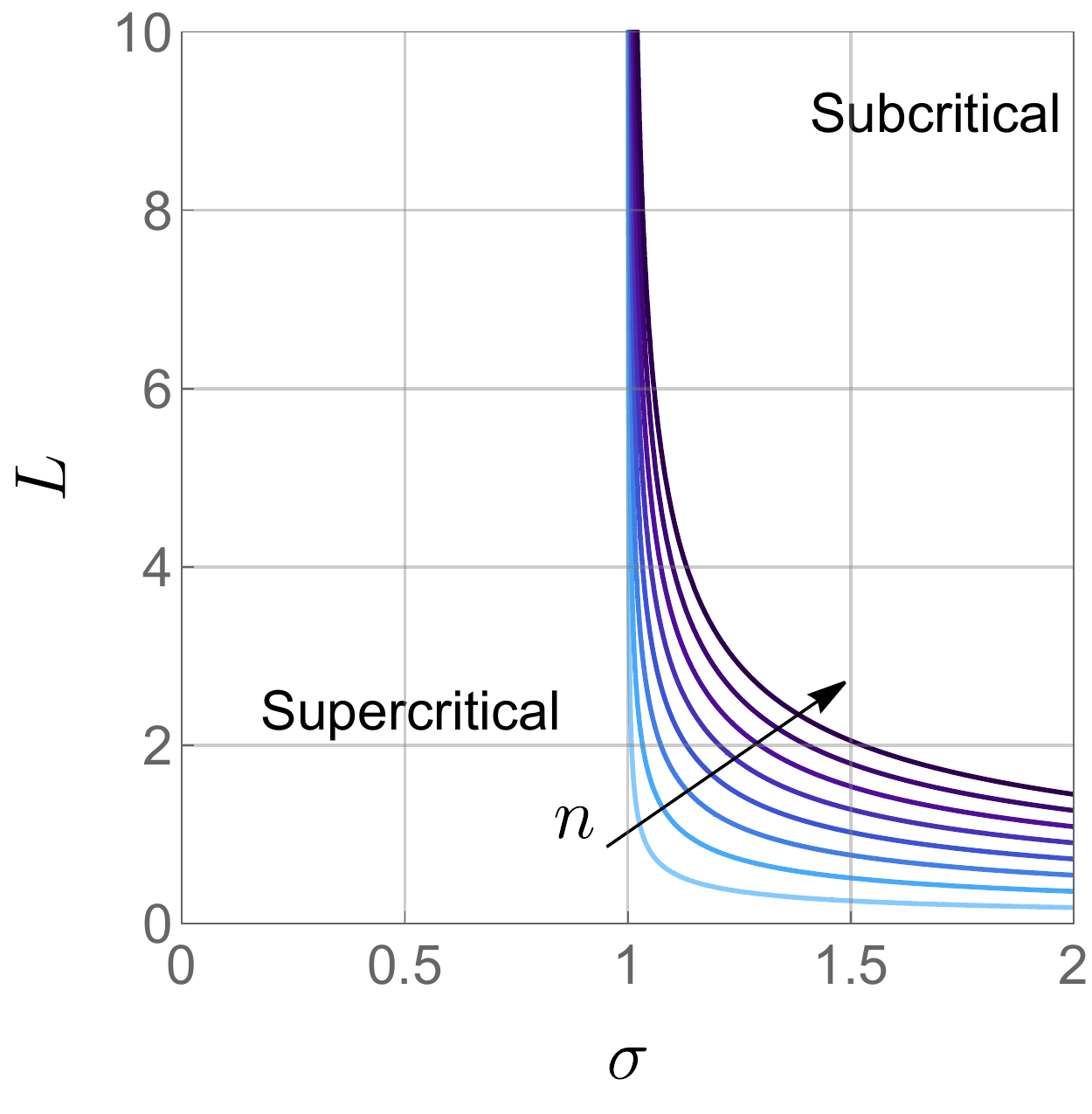}\includegraphics[width=0.33\textwidth]{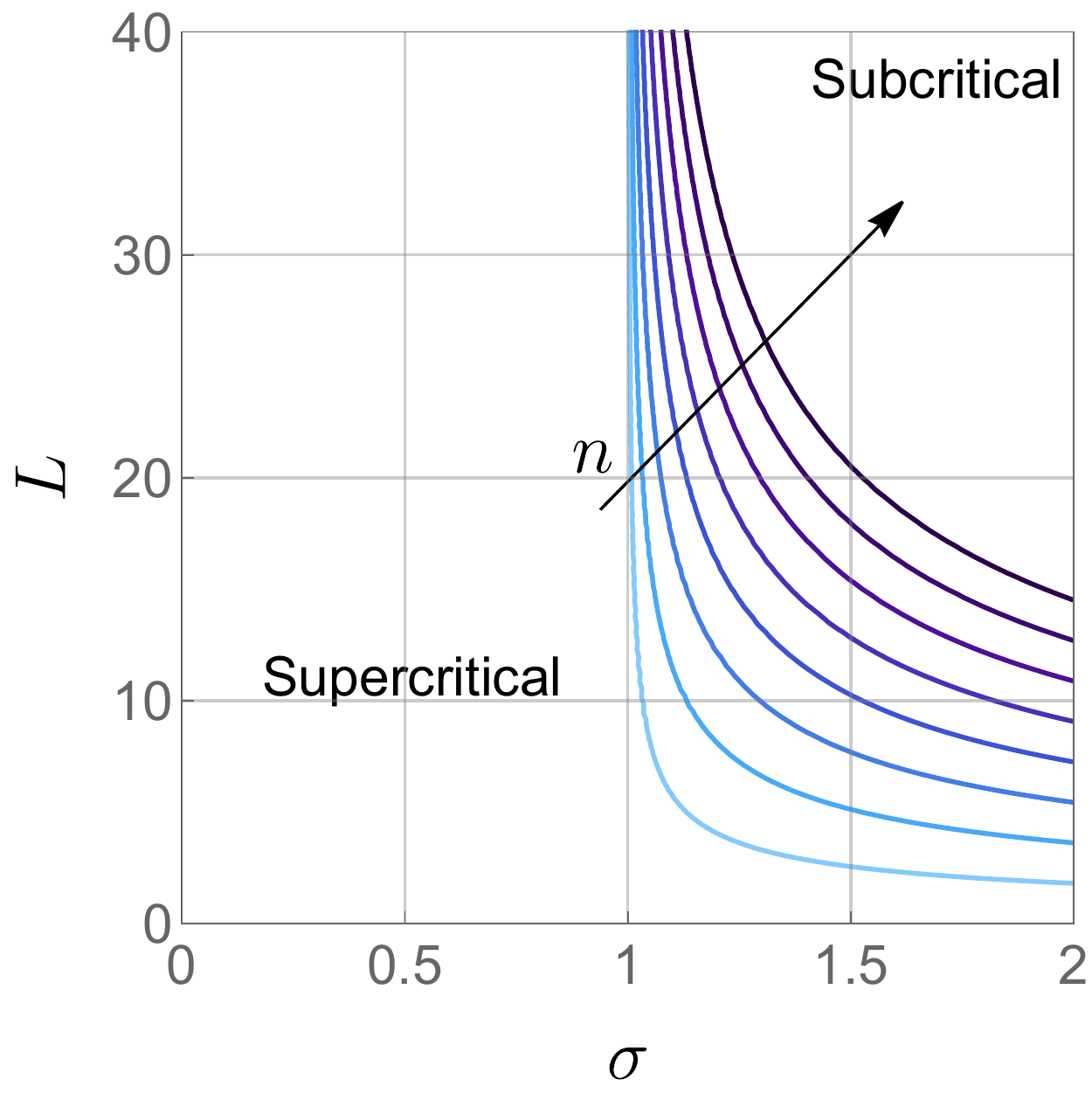}\includegraphics[width=0.33\textwidth]{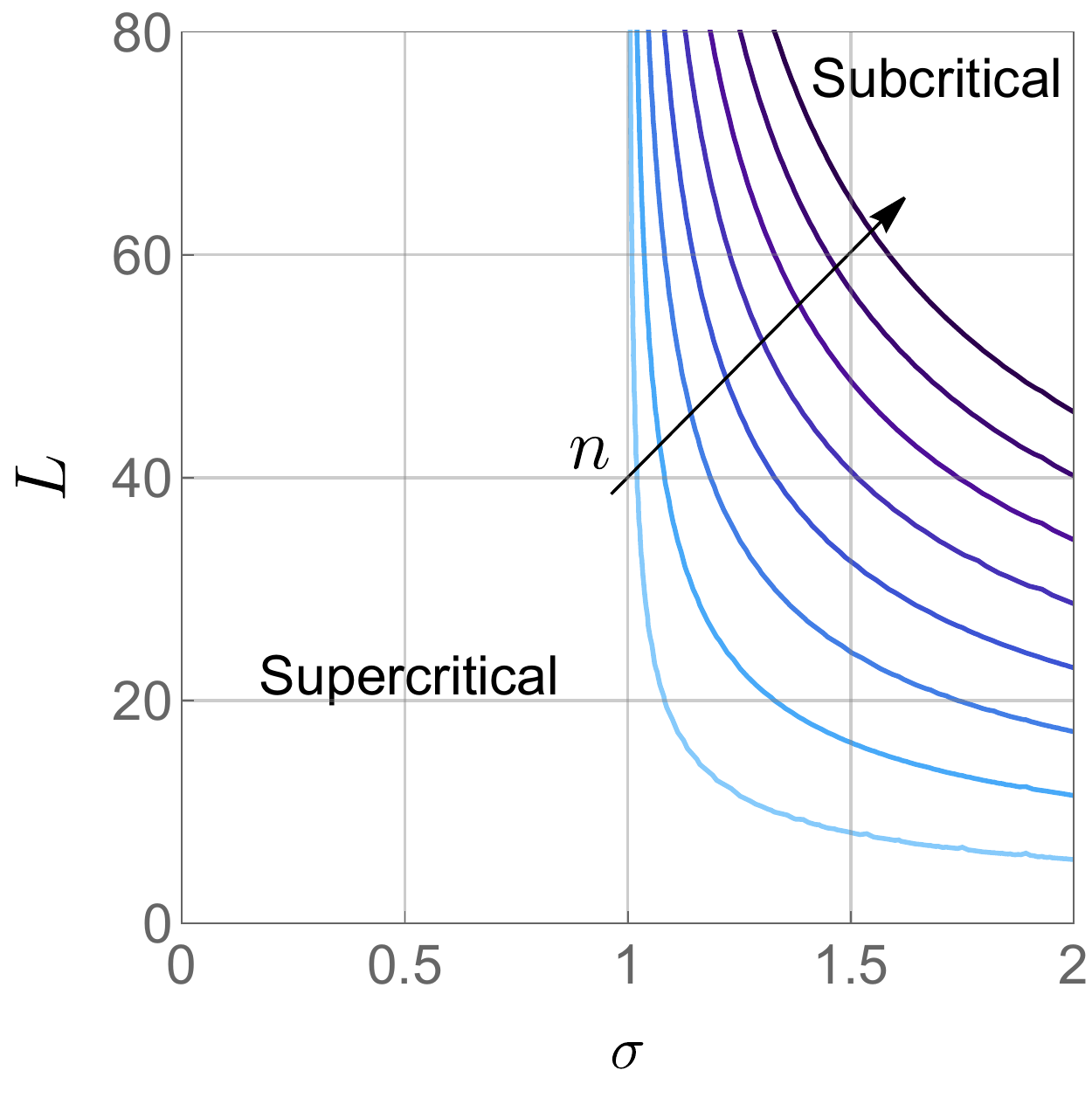}
\caption{Phase diagram showing the transition from a supercritical to a subcritical pitchfork bifurcation for $\beta = 0.01,\,1,\,10$ and $u_1^{\star}=u_3^{\star}=0$ (left, center and right respectively) where the control parameter $\lambda=u_2^{\star}$ and $F=0$. The lines corresponds to the wavenumber $n=1,\,\dots,\,8$, where the arrows denote the direction in which $n$ increases.}
\label{fig:sub_super_trans}
\end{figure}

We can now compute $\lambda_2$: setting equal to zero the coefficient multiplying $\xi^3$ in \eqref{eq:expansion_energy} and $\delta\omega=\omega_1$, we get
\begin{equation}
\label{eq:lambda_2_budiansky}
\lambda_2 = -\frac{\frac{1}{6}\delta^4\Psi_c[\omega_1,\,\omega_1,\,\omega_1,\,\omega_1]+\delta^3\Psi_c[\omega_1,\,\omega_1,\,\omega_2]}{\delta^2\dot\Psi_c[\omega_1,\,\omega_1]}.
\end{equation}
As observed before, we have that $\delta^3\Psi_c[\omega_1,\,\omega_1,\,\omega_2]=0$. From the expression of the energy given by~\eqref{eq:ener_omega} and from \eqref{eq:omega_1_perversion}, we obtain
\begin{equation}
\label{eq:lambda_2}
\lambda_2 = \frac{1}{4} \left(\frac{\pi ^4 \beta  n^4}{L^4}-\frac{3 \pi ^2 n^2 (\sigma -1)}{L^2}\right).
\end{equation}

It is now possible to characterize the bifurcation near the marginal stability threshold: neglecting the higher order terms in \eqref{eq:expansion_v_u_1^*} and since $\lambda_1$ vanishes, we get
\[
\xi = \pm\sqrt{\frac{u_2^{\star}-\lambda_0}{\lambda_2}}.
\]
Since $\lambda_1=0$ and $\lambda_2\neq0$, we are in the presence of a pitchfork bifurcation, a supercritical or subcritical one depending on whether $\lambda_2$ is positive or negative, respectively (see figure~\ref{fig:sub_super_trans}).
\subsubsection{Weakly nonlinear analysis in the presence of an axial load}

In this section, we perform again the weakly nonlinear analysis using as control parameter of the problem $\lambda=-F$. Since the procedure is analogous to that of the previous section, we report the main results omitting the algebraic computations. The critical load at which the straight configuration bifurcates into a shape with $n$ perversions is given by \eqref{eq:sol_pin_end}, so that
\[
\lambda_0 = \frac{\pi ^2 \beta  n^2}{L^2}+\sigma-2u_2^{\star},
\]
where the mode $\omega_1$ is given by \eqref{eq:omega_1_perversion}. Following the same steps as in section~\ref{sec:WNL_u2s}, we get that $\lambda_1 = 0$, see \eqref{eq:lambda_1}. 

\begin{figure}[t!]
\includegraphics[width=0.33\textwidth]{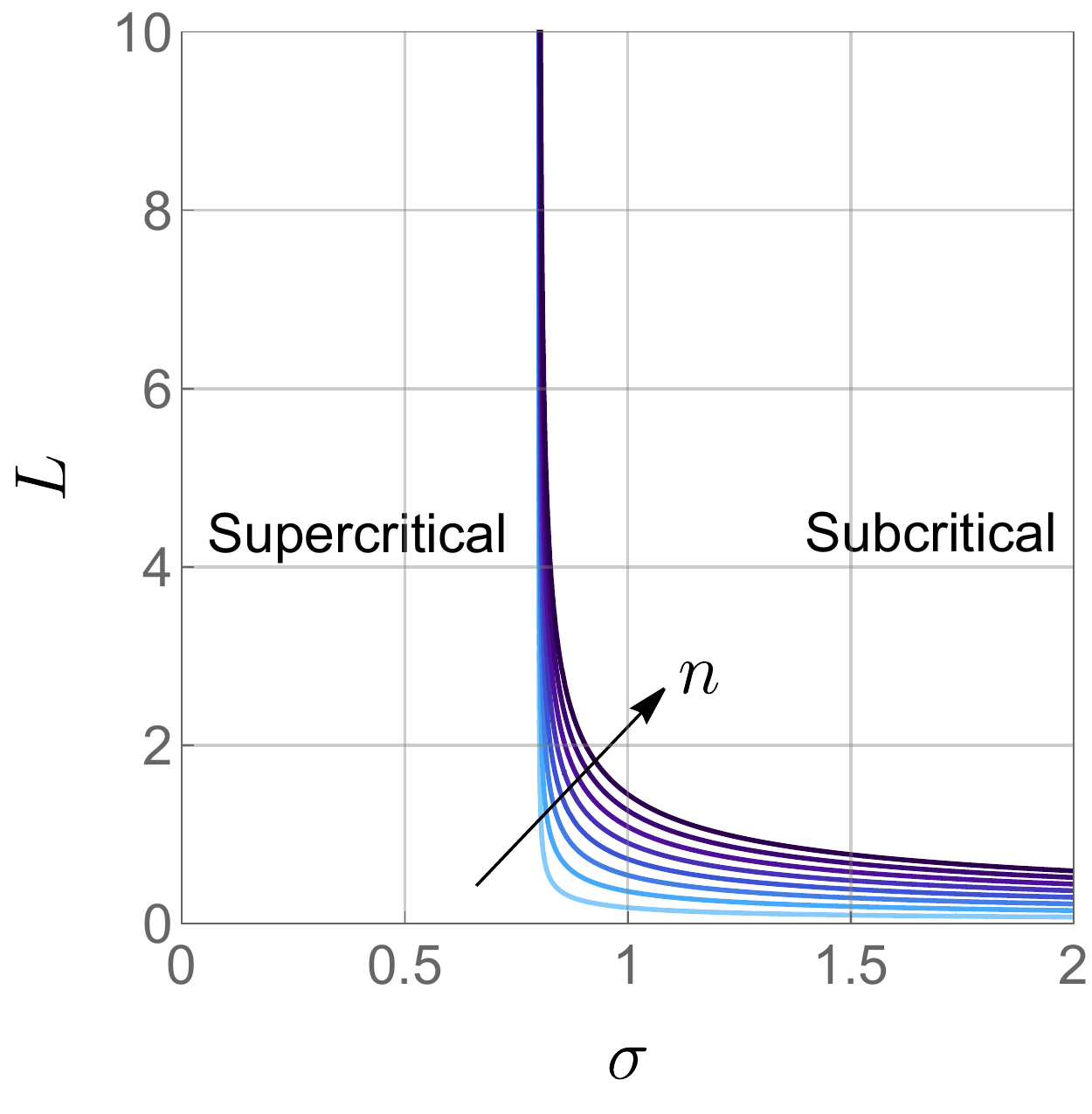}\includegraphics[width=0.33\textwidth]{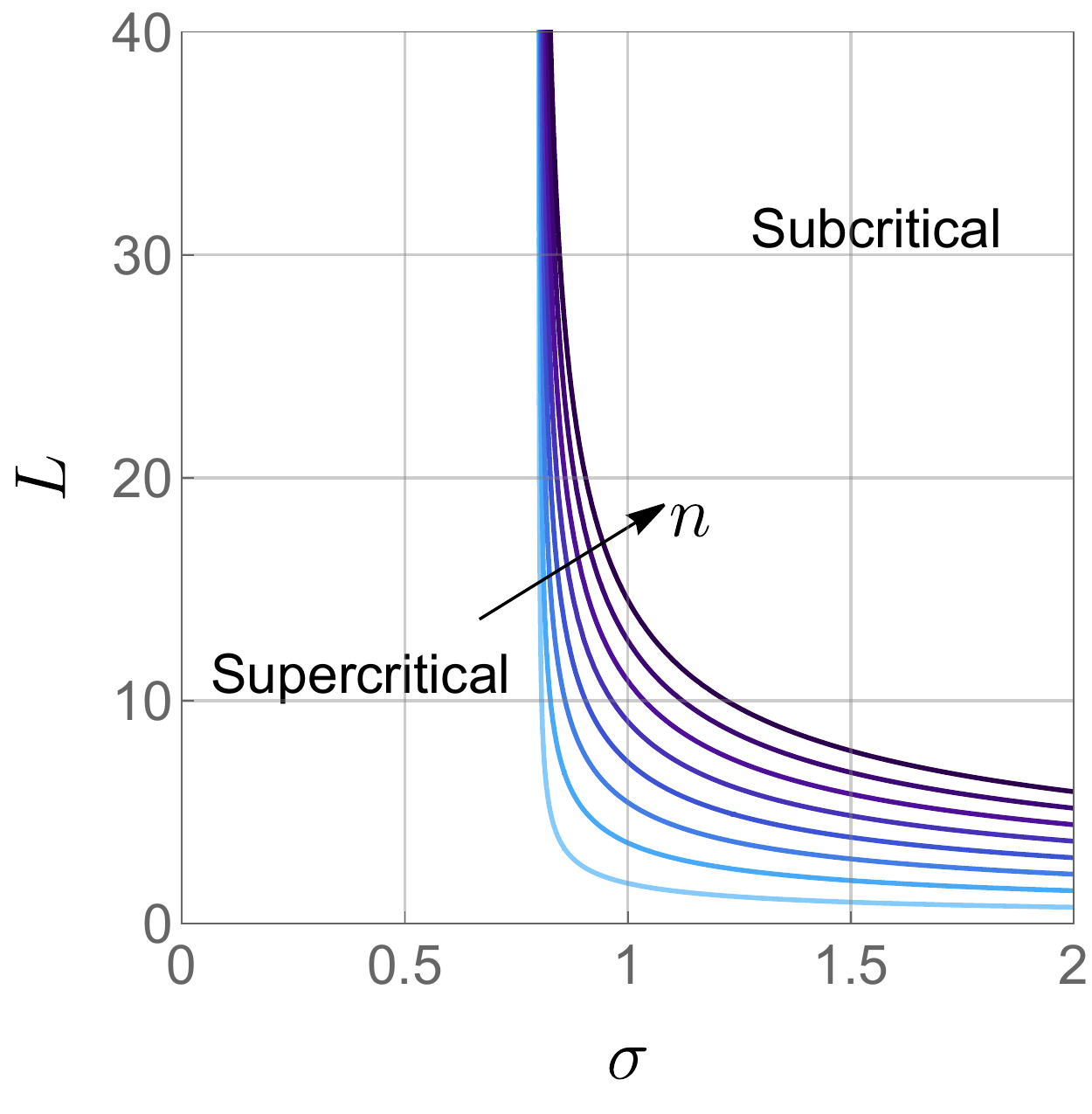}\includegraphics[width=0.33\textwidth]{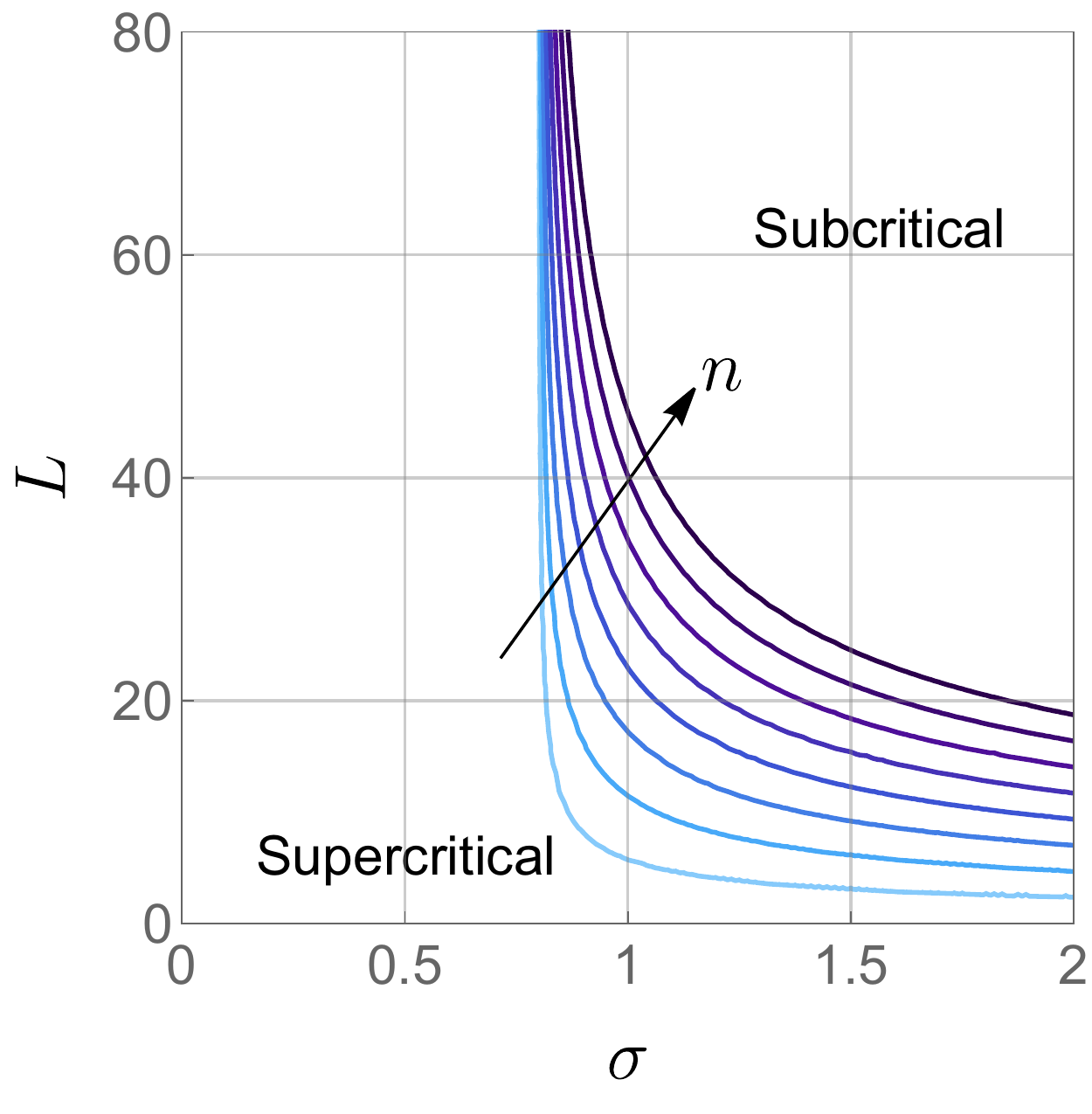}
\caption{Phase diagram showing the transition from supercritical to a subcritical pitchfork bifurcation for $\beta = 0.01,\,1,\,10$ (left, center and right respectively) and $u_1^{\star}=u_2^{\star}=u_3^{\star}=0$,  where the control parameter is $\lambda=-F$. The lines corresponds to the wavenumber $n=1,\,\dots,\,8$, where the arrows denote the direction in which $n$ increases.}
\label{fig:sub_super_trans_F}
\end{figure}

Finally, to compute $\lambda_2$, we use again the formula \eqref{eq:lambda_2_budiansky}, obtaining
\begin{equation}
\label{eq:l2_F}
\lambda_2=\frac{1}{8} \left(\frac{\pi ^4 \beta  n^4}{L^4}+\frac{3 \pi ^2 n^2 (-5 \sigma +2 u_2^{\star}+4)}{L^2}\right).
\end{equation}

From these formulae, neglecting the higher order terms, we can express the amplitude $\xi$ of the buckled configuration as
\begin{equation}
\label{eq:xi_F}
\xi = \pm\sqrt{\frac{-F-\lambda_0}{\lambda_2}}.
\end{equation}
From \eqref{eq:xi_F}, we can observe that the instability corresponds to a pitchfork bifurcation, since the amplitude grows as the square root of the distance from the marginal stability threshold, and it is supercritical or subcritical if the sign of $\lambda_2$ is positive or negative, respectively (see figure~\ref{fig:sub_super_trans_F}).

\subsection{Differences in the pinned ends case}
If we consider the pinned ends case, the theoretical analysis performed in the previous sections can be easily adapted to these boundary conditions. 
All the helical configurations are ruled out by the homogeneous boundary conditions $\omega(0) = \omega(L) = 0$. By enforcing the natural boundary conditions \eqref{eq:BC_caso_B}, $\omega = 0$ is still a solution of the problem when $u_1^{\star}=0$. The main difference with respect to case A is that now the reference state is always an equilibrium configuration irrespective of $u_3^\star$. This fact does not affect the linearized problem, as shown by \eqref{eq:ELlinear}, and the critical buckling mode corresponds to the wavenumber $n=1$ in \eqref{eq:omega_1_perversion}. 
Some caution is needed while carrying out the weakly nonlinear analysis: since $u_3^\star$ can be non-zero, we get
\[
\delta^3\Psi_c[v_1,\,v_2,\,v_3] = 3 \int_0^L\sigma  u_3^\star v_1'(s) v_2'(s)v_3'(s)\,ds.
\]
As $\delta^3\Psi_c[\omega_1,\,\omega_1,\,\omega_1]=0$, then $\lambda_1 = 0$ for both the control parameters $u_2^\star$ and $F$ (see  see \eqref{eq:lambda_1}). However, to compute $\lambda_2$ in the expansion \eqref{eq:expansion_v_u_1^*}, we need to calculate the second order term $\omega_2$. Setting the second order term of \eqref{eq:expansion_energy} equal to zero for a general perturbation $\delta\omega$, we get the following boundary value problem
\[
\left\{
\begin{aligned}
&\beta  \omega_2''''(s)+\frac{\pi ^2 \beta  n^2 \omega_2''(s)}{L^2}+\frac{3 \pi ^3 n^3 \sigma  u_3^{\star} \sin \left(\frac{2 \pi  n s}{L}\right)}{2 L^3}=0,\\
&\omega_2(0)=\omega_2(L)=\omega_2''(0)=\omega_2''(L)=0.
\end{aligned}
\right.
\]
A solution of this problem, satisfying the orthogonality condition \eqref{eq:orthogonality}, is given by
\begin{equation}
\label{eq:omega_2}
\omega_2(s) = -\frac{L \sigma  u_3^{\star} }{8 \pi  \beta  n}\sin \left(\frac{2 \pi  n s}{L}\right).
\end{equation}
Summing up, using \eqref{eq:lambda_2_budiansky}, it is possible to show that, if $\lambda=u_2^\star$, then
\[
\lambda_2 =  \frac{1}{16} \left(\frac{4 \pi ^4 \beta  n^4}{L^4}-\frac{12 \pi ^2 n^2 (\sigma -1)}{L^2}-\frac{3 \sigma ^2 {u_3^{\star}}^2}{\beta }\right);
\]
while if the control parameter is the axial force $F$, then
\[
\lambda_2 = \frac{1}{8} \left(\frac{\pi ^4 \beta  n^4}{L^4}+\frac{3 \pi ^2 n^2 (-5 \sigma +2 u_2^{\star}+4)}{L^2}-\frac{3 \sigma ^2 {u_3^{\star}}^2}{\beta }\right).
\]

In the following, we complement the theoretical analysis performed in the previous sections with numerical simulations together with a comparison with some experiments.

\section{Numerical simulations and experimental results}
\label{sec:numerics}
In this section, we present the results of numerical simulations based on a finite element approximation of the nonlinear problem. The numerical outcomes are compared with the theoretical predictions of the previous sections as well as with some experimental realizations of the mechanical system explored in this paper.
The technical details on the numerical algorithm and of the experimental setup are reported in the electronic supplementary material, while the numerical code is available in a GitHub repository whose link is reported in the Data Accessibility statement.

\subsection{Case A: Free ends}
\begin{figure}[t!]
\begin{center}
\begin{minipage}{0.5\textwidth}
\begin{center}
\includegraphics[width=\textwidth]{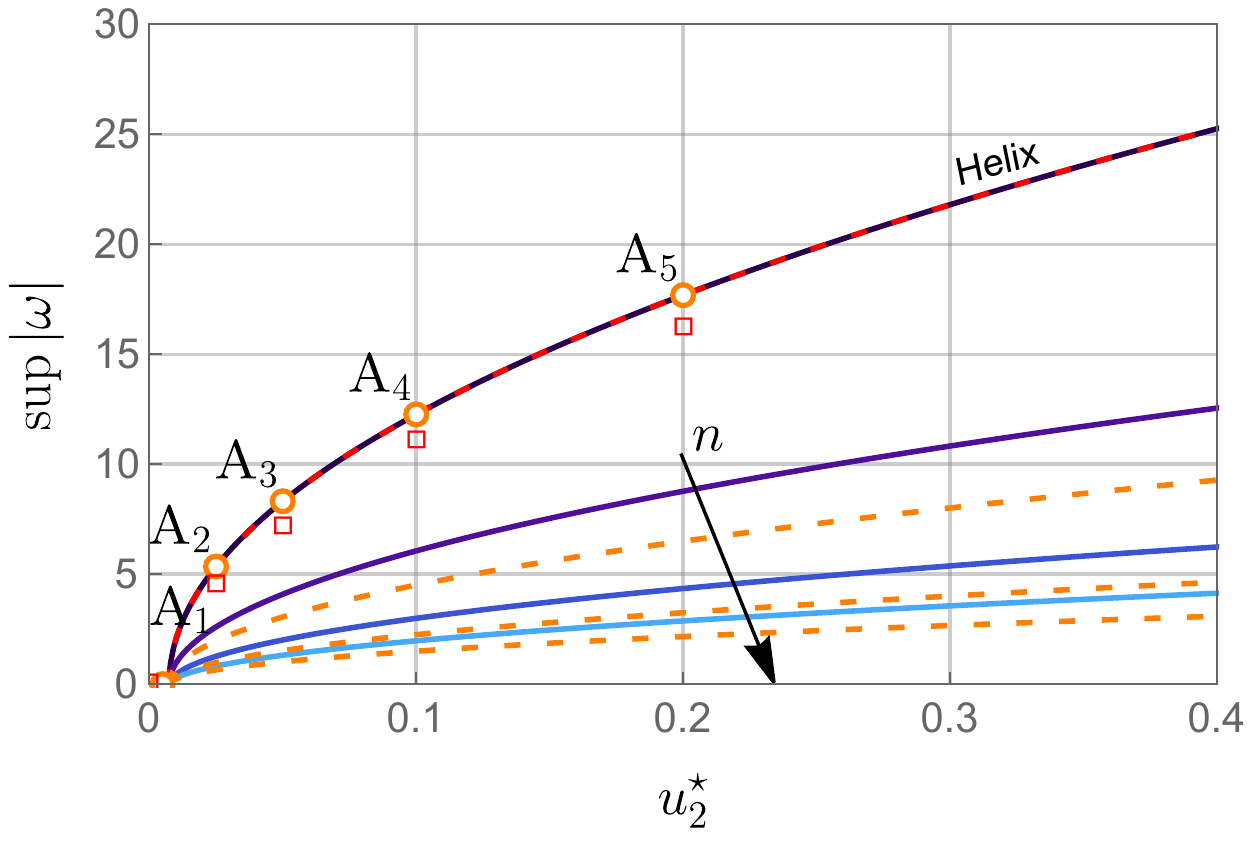}\\
\includegraphics[width=\textwidth]{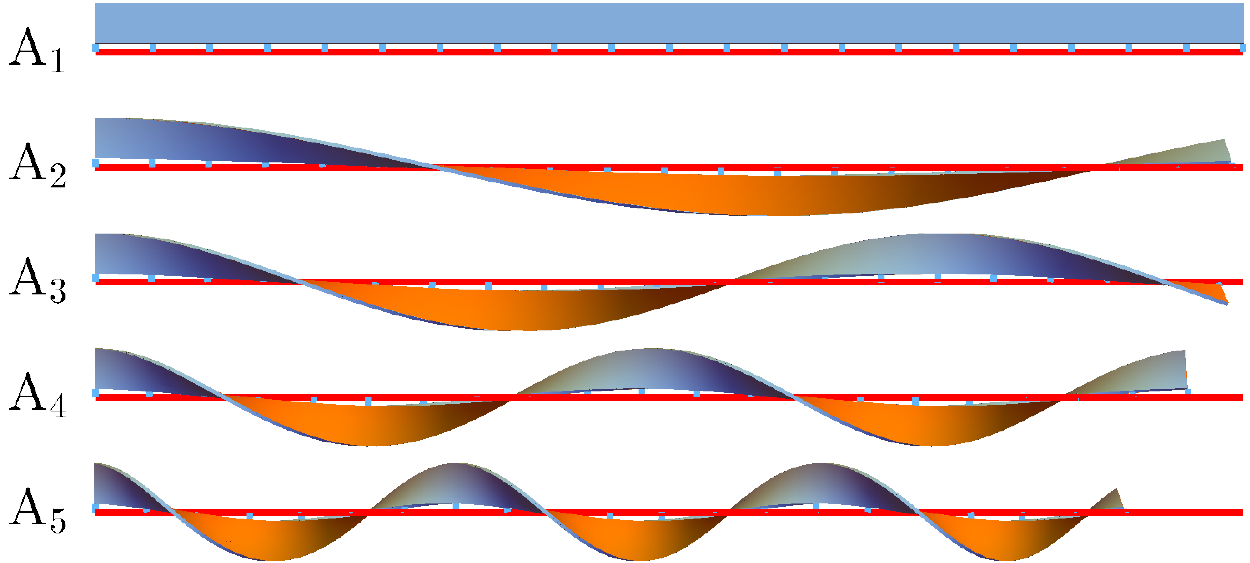}
\end{center}
\end{minipage}\begin{minipage}{0.5\textwidth}
\begin{center}
\includegraphics[width=\textwidth]{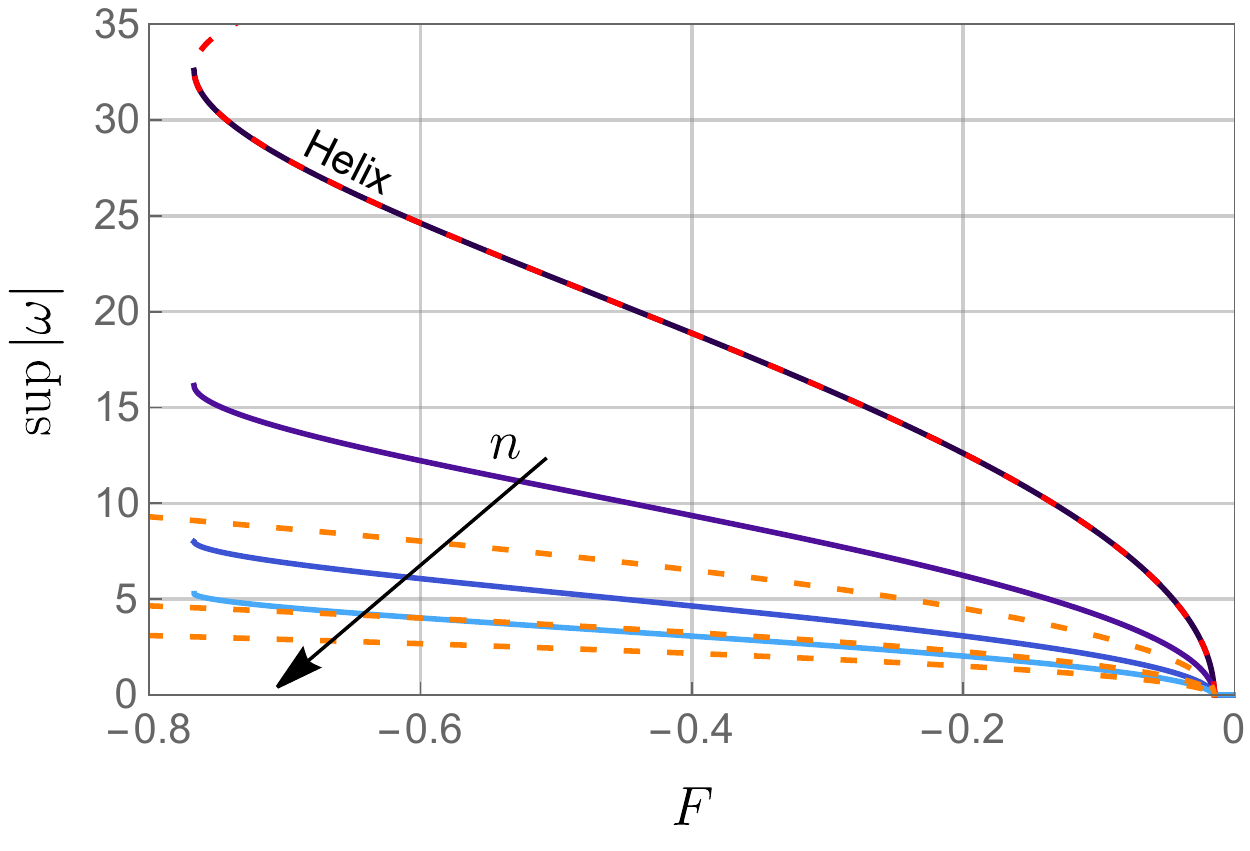}\\
\includegraphics[width=0.95\textwidth]{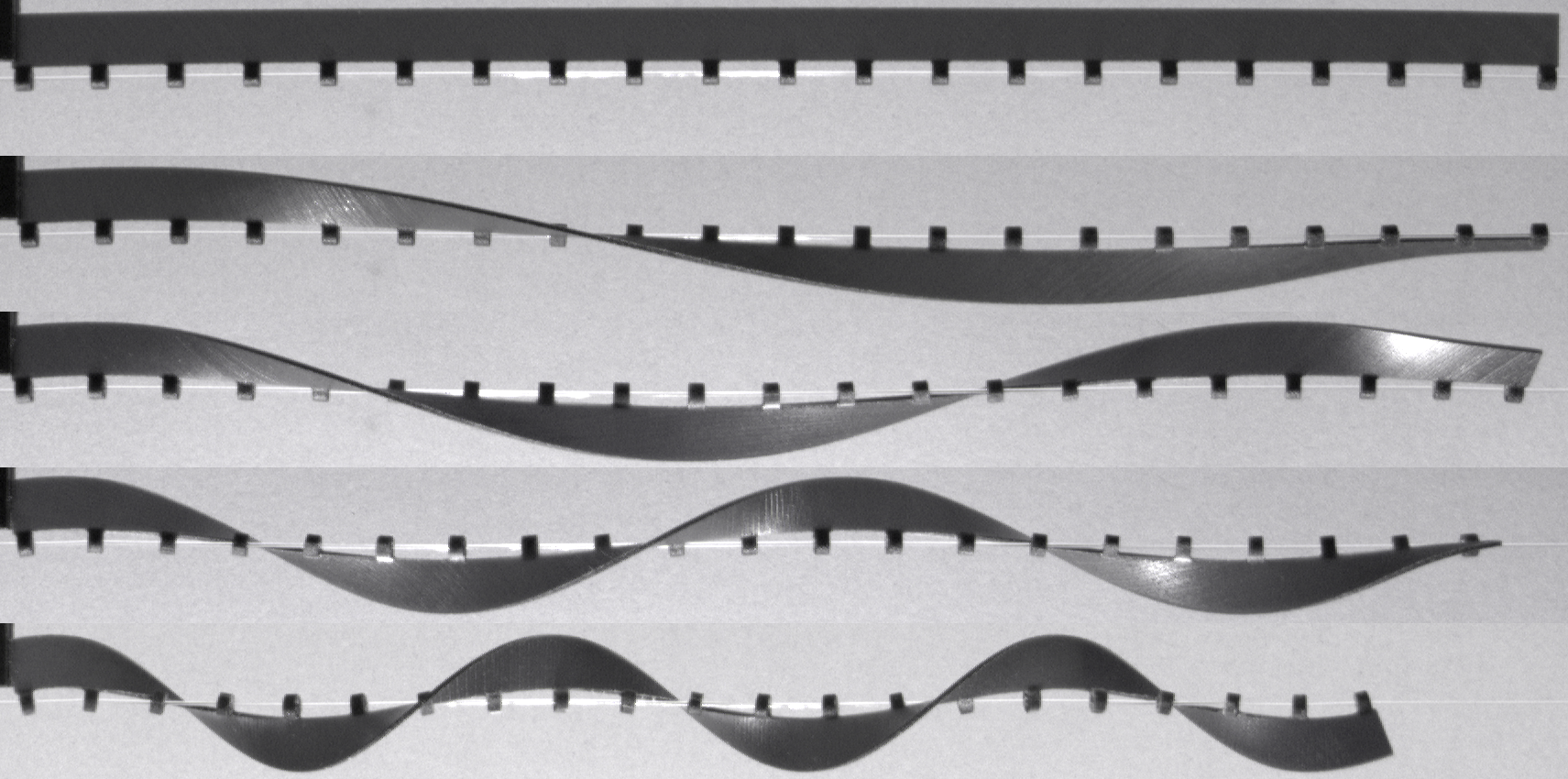}%
\end{center}
\end{minipage}
\end{center}
\caption{(Top) Bifurcation diagrams obtained by using $u_2^{\star}$ (left) and $F$ (right) as control parameters for a rod having a rectangular cross-section ($h/t=10$, $L=40$) in the free end case. The dashed lines in the bifurcation diagrams represent the analytical helical configuration and the weakly nonlinear analysis predictions (red and orange lines respectively). The circles and squares represent the numerical and experimental results, respectively, corresponding to the snapshots shown in the bottom part. (Bottom) Comparison of numerical (left) and experimental (right) post-buckling evolution of the critical mode when the bifurcation is triggered by the natural curvature $u_2^\star$. The numerical and experimental configurations in the bottom panels are obtained for increasing values of $u_2^*$, ranging from $0.005$ for $\mathrm{A}_1$ to $0.2$ for $\mathrm{A}_5$. The Poisson's ratio has been set equal to $0.35$, while $\chi=1$ in \eqref{eq:betasigmarect}}
\label{fig:num_res_free_t_01}
\end{figure}

We first present the results in the case of free ends. In figure~\ref{fig:num_res_free_t_01}, we show the outcomes for a rod having a rectangular cross-section whose aspect ratio is $h/t=10$ and $L=40$. We have performed two simulations. In the first one we have used $u_2^{\star}$ as control parameter, setting $u_1^{\star}=u_3^{\star}=F=0$. In the latter, the control parameter was $-F$ and all the natural curvatures and twist were set equal to zero.
In the top panels we plot the bifurcation diagrams, where $\sup |\omega|$ is used as a measure of the amplitude of the buckled solution. The numerical results perfectly match with the analytical solution of the helical configuration, represented by the red dashed line, see \eqref{eq:F_alpha}, validating the numerical code. In the neighborhood of the marginal stability thresholds, there is a good agreement between the numerical simulations and the results of the weakly nonlinear analysis.

Frequently, rods with a high cross-sectional aspect ratio are modeled using a ribbon model, based on the Wunderlich's functional \cite{Wunderlich_1962,dias2016wunderlich}. Indeed, as $h\gg t$, this model shows a better agreement with experimental results compared with the Kirchhoff's rod model exploited in this paper \cite{kumar2020investigation}. These discrepancies can be related to the aspect ratio of the cross-section \cite{moulton2018stable}, as well as to the kind of deformation the rods are subject to \cite{yu2019bifurcations}. However, as shown figure~\ref{fig:num_res_free_t_01}, we find a good quantitative agreement between experimental results (reported on the top left plot as the red squares) and the model predictions based on the Kirchhoff's energy functional \eqref{eq:energy_one_rod}. This provides evidence of the validity of our constitutive choice for the considered range of the parameters. This is in agreement with the results of \cite{moulton2018stable}, where the authors show the validity of the Kirchhoff's model when $h/t=O(10)$. In figure~\ref{fig:num_res_free_t_01} (bottom panels) we also report the post-buckling configurations predicted by the numerical simulations side by side with the pictures taken during the experiments. 

\subsection{Case B: Pinned ends}
\begin{figure}[t!]
\begin{center}
\begin{minipage}{0.5\textwidth}
\begin{center}
\includegraphics[width=\textwidth]{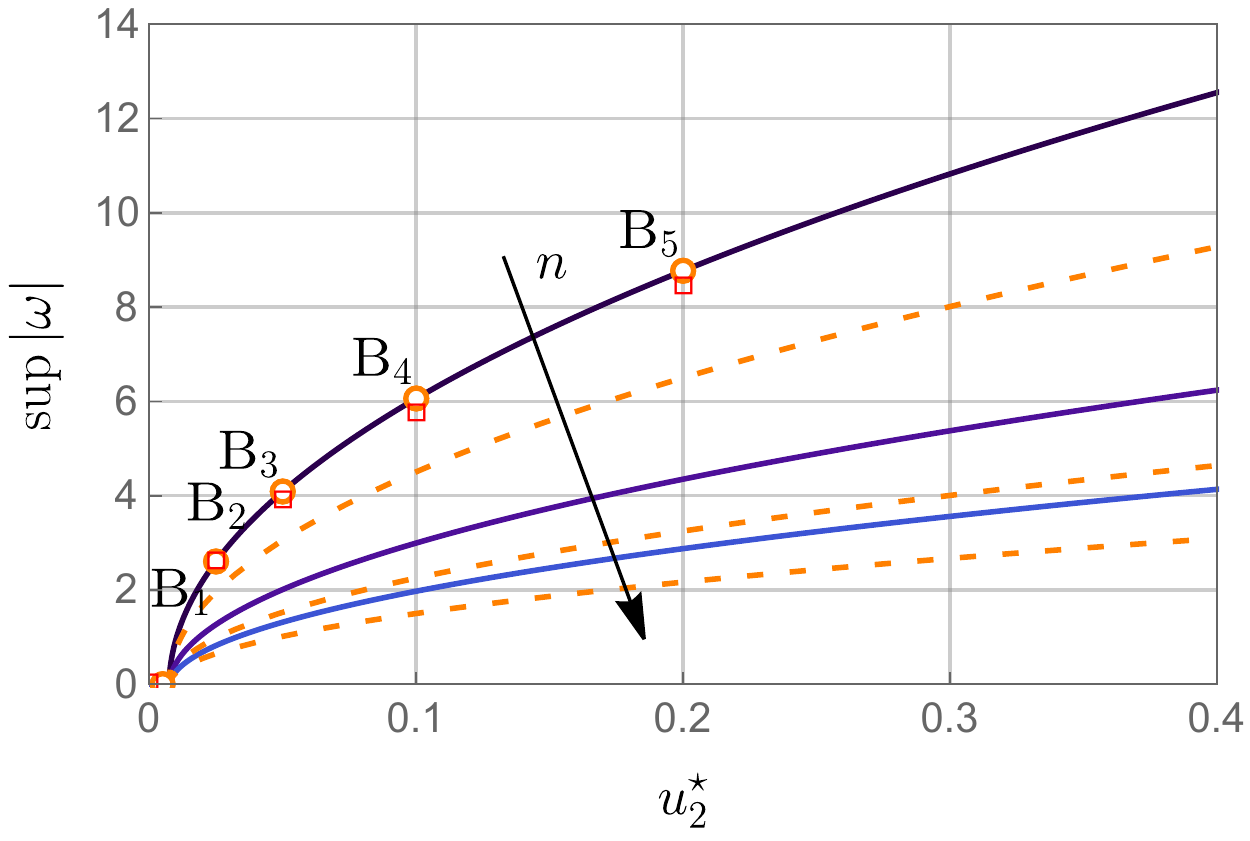}\\
\includegraphics[width=\textwidth]{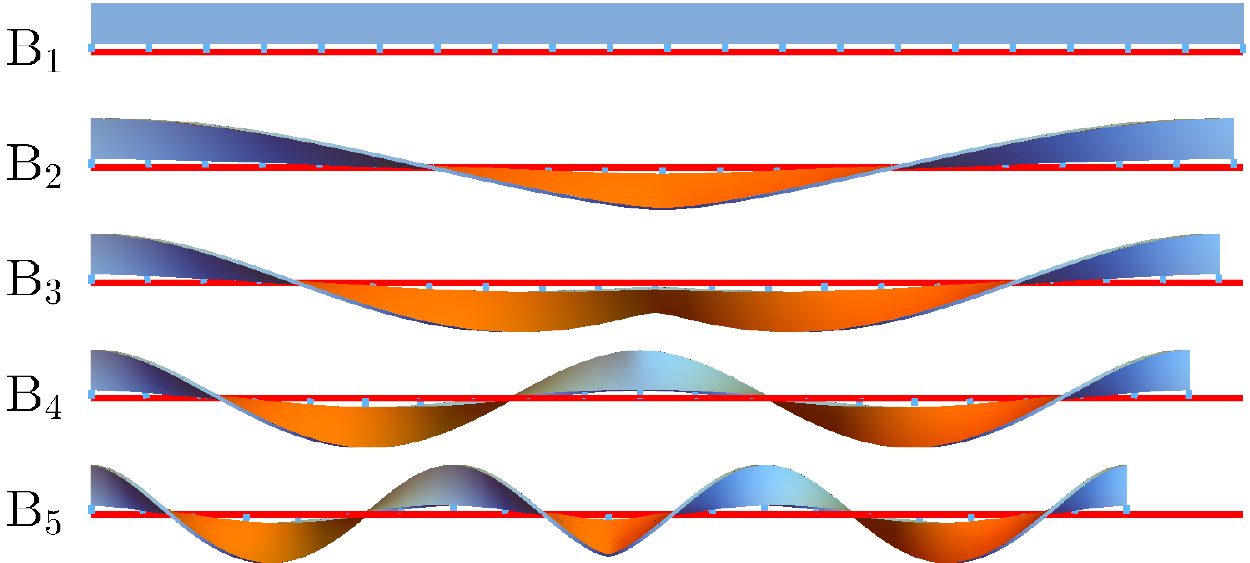}
\end{center}
\end{minipage}\begin{minipage}{0.5\textwidth}
\begin{center}
\includegraphics[width=\textwidth]{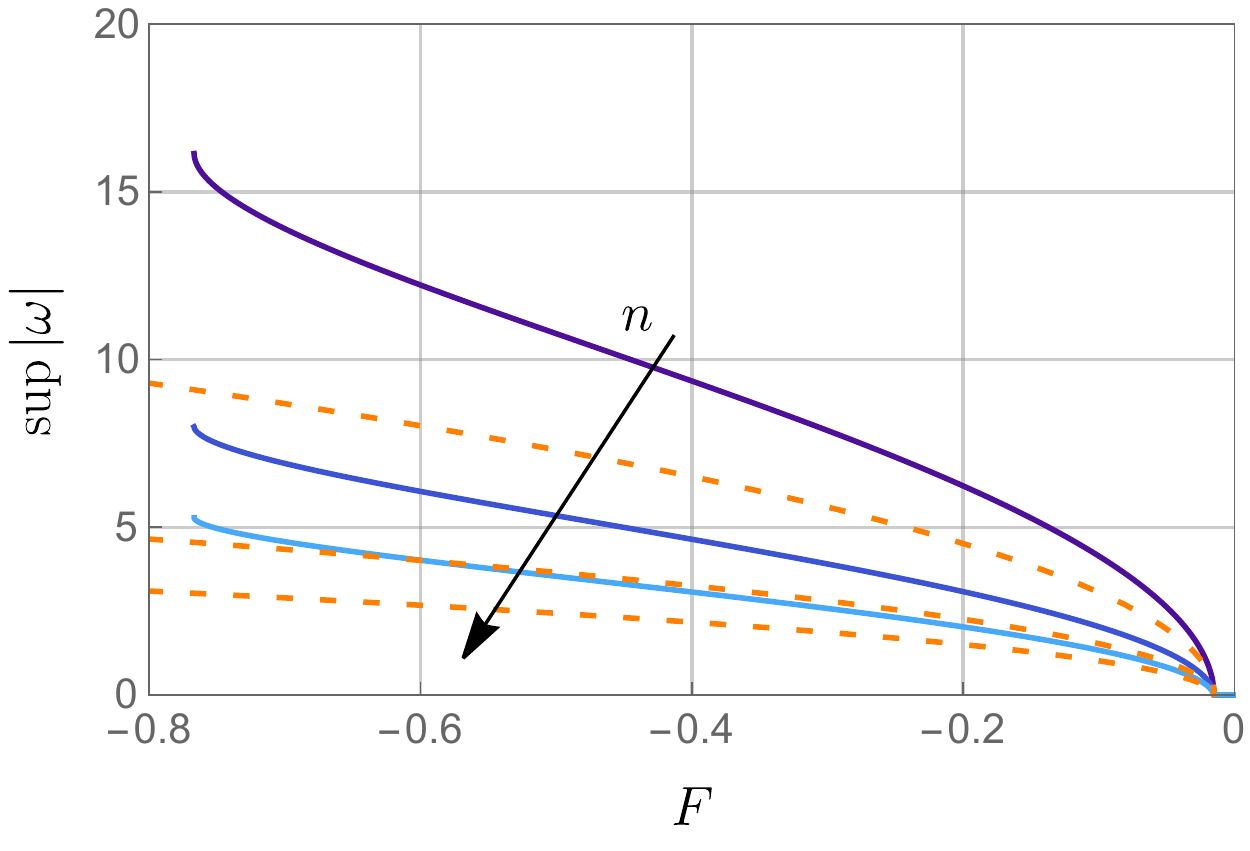}\\
\includegraphics[width=0.95\textwidth]{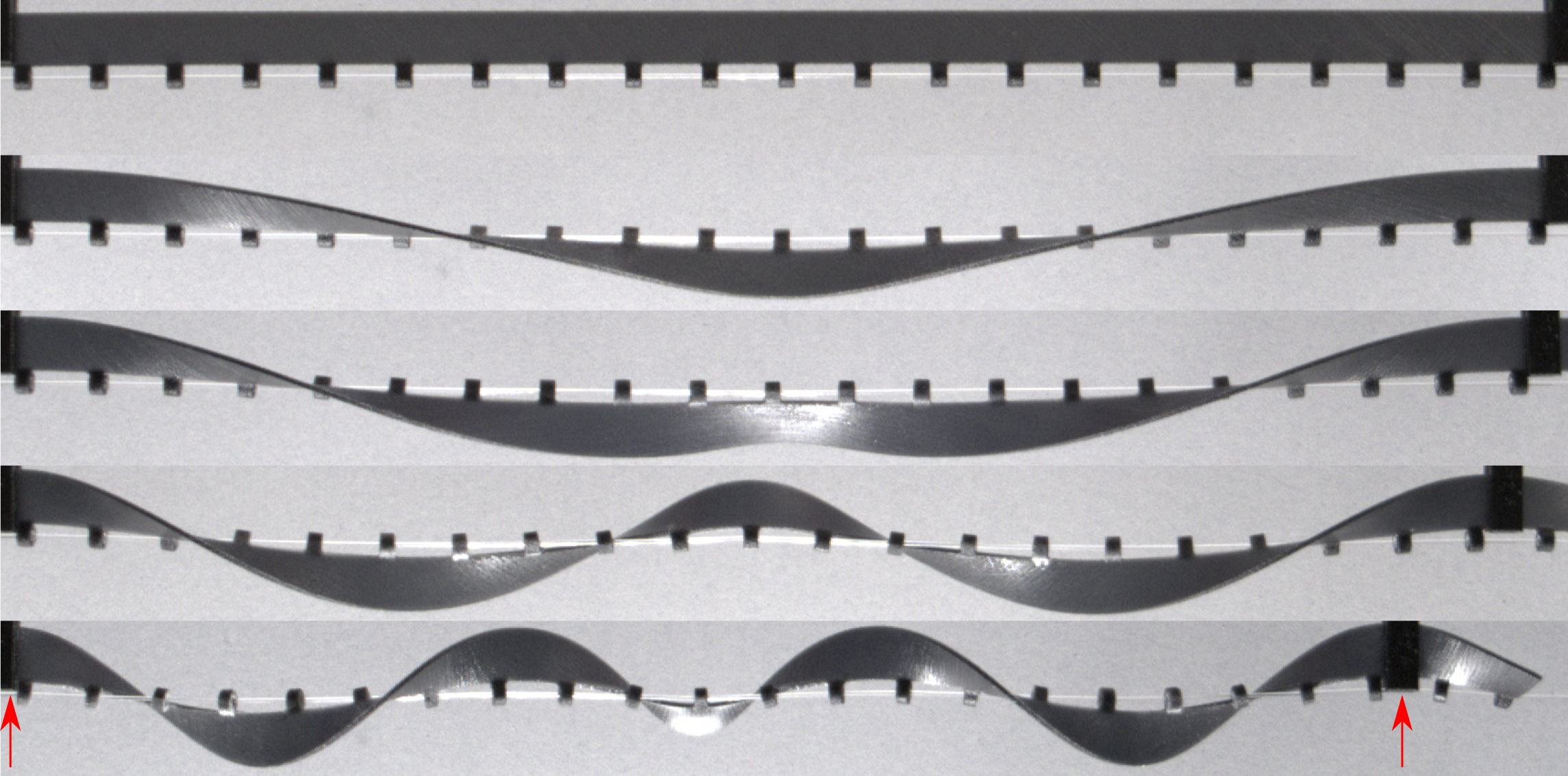}
\end{center}
\end{minipage}
\end{center}
\caption{(Top) Bifurcation diagrams obtained by using $u_2^{\star}$ (left) and $F$ (right) as control parameters for a rod having a rectangular cross-section ($h/t=10$, $L=40$) in the pinned ends case.
The dashed lines in the bifurcation diagrams represent the weakly nonlinear analysis predictions (orange lines). The circles and squares represent the numerical and experimental results, respectively, corresponding to the snapshots shown in the bottom part. (Bottom) Comparison of numerical (left) and experimental (right) post-buckling evolution of the critical mode when the bifurcation is triggered by the natural curvature $u_2^\star$. The numerical and experimental configurations in the bottom panels are obtained for increasing values of $u_2^*$, ranging from $0.005$ for $\mathrm{B}_1$ to $0.2$ for $\mathrm{B}_5$. In the experiments, the boundary conditions $\omega(0)=\omega(L)=0$ are enforced by means of 3D printed supports preventing the rotation of the rod extremities (indicated with red arrows in the experimental realization of $\mathrm{B}_5$). The Poisson's ratio has been set equal to $0.35$, while $\chi=1$ in \eqref{eq:betasigmarect}}
\label{fig:num_res_pinned_t_01}
\end{figure}
\begin{figure}[t!]
\centering
\includegraphics[width=0.5\textwidth]{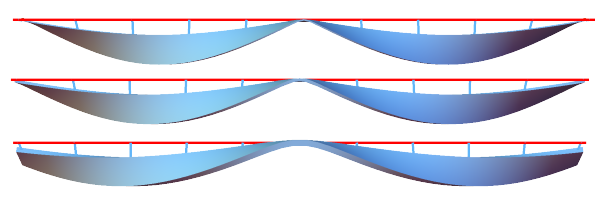}%
\includegraphics[width=0.5\textwidth]{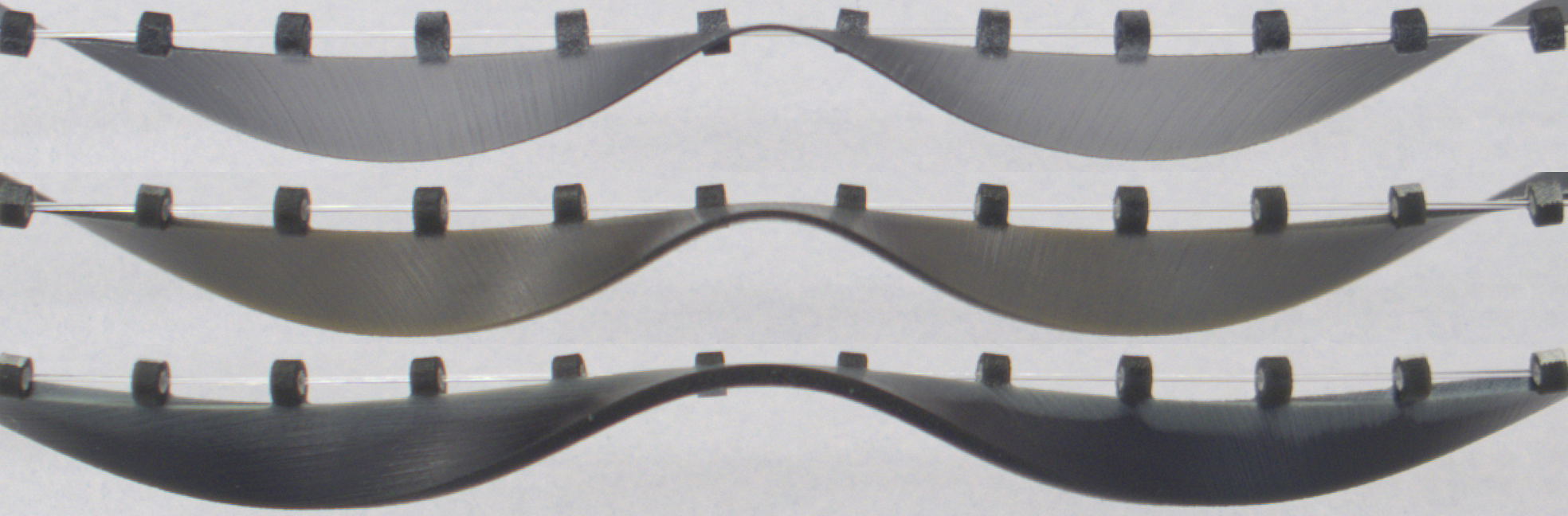}
\caption{Detail of the transition zone (perversion) exhibited by a rod having $L=40$, $u_2^*=0.1$, and $h/t=20,\,10,\,5$ (top, center, bottom). We observe that as $h/t$ grows, the perversion is more localized.}
\label{fig:loc}
\end{figure}
The outcomes of the numerical simulations are shown in figure~\ref{fig:num_res_pinned_t_01}.
Not surprisingly, the bifurcation diagrams shown in figure \ref{fig:num_res_pinned_t_01} are identical to those shown in figure~\ref{fig:num_res_free_t_01}, except for the branch corresponding to helical configurations. In fact, these solutions are ruled out by the boundary conditions $\omega(0)=\omega(L)=0$. 
This implies that the critical mode is the one corresponding to a configuration exhibiting a single perversion: in the post-buckling regime, the shape of the rod evolves towards two helices having opposite chirality, connected by a transition zone.
As for Case A, we obtain a good quantitative agreement with the experimental results, see figure~\ref{fig:num_res_pinned_t_01} (reported on the top left plot as the red squares). We also report the post-buckling configurations predicted by the numerical simulations side by side with the pictures taken during the experiments (figure~\ref{fig:num_res_pinned_t_01}, bottom panels).

Interestingly, the length of the transition zone appears to be correlated with the cross section of the rod. As shown in figure~\ref{fig:loc}, thinner rods exhibit a localization of the perversion.
The good agreement between numerical computations and experiments proves that the Kirchhoff's functional \eqref{eq:energy_one_rod} is an appropriate constitutive choice for the considered range of the parameters, just as for the previous case.


\subsection{Competition between helix and modes exhibiting a single or multiple perversions}
\label{sec:numerics_f}

\begin{figure}[t!]
\includegraphics[width=\textwidth]{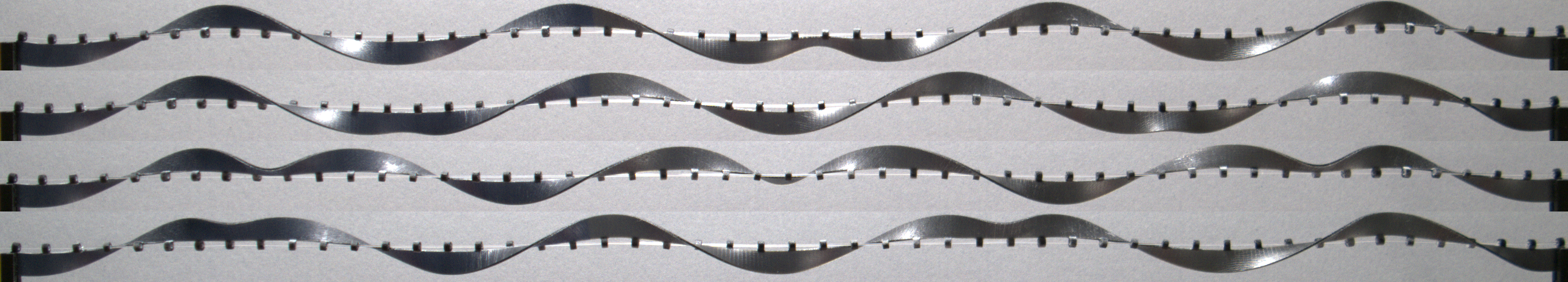}
\caption{Different equilibrium configuration obtained for an experimental rod with $L=100$, $h/t=10$, $u_2^*=0.1$, exhibiting $1$ (top)$,\,2,\,3,\,4$ (bottom) perversions.}
\label{fig:multiple_perv}
\end{figure}

The results of the previous sections show that the critical buckling modes are characterized by either helical shapes or morphologies exhibiting a single perversion, depending on the boundary conditions applied to the rod.
However, during our experiments, we have observed a more complex multi-stable behavior: helices and single or multiple perversions can be experimentally observed by manually deforming the rod into such shapes, that seem to be stable (see figure~\ref{fig:multiple_perv}).

\begin{figure}
\centering
\includegraphics[width=0.5\textwidth]{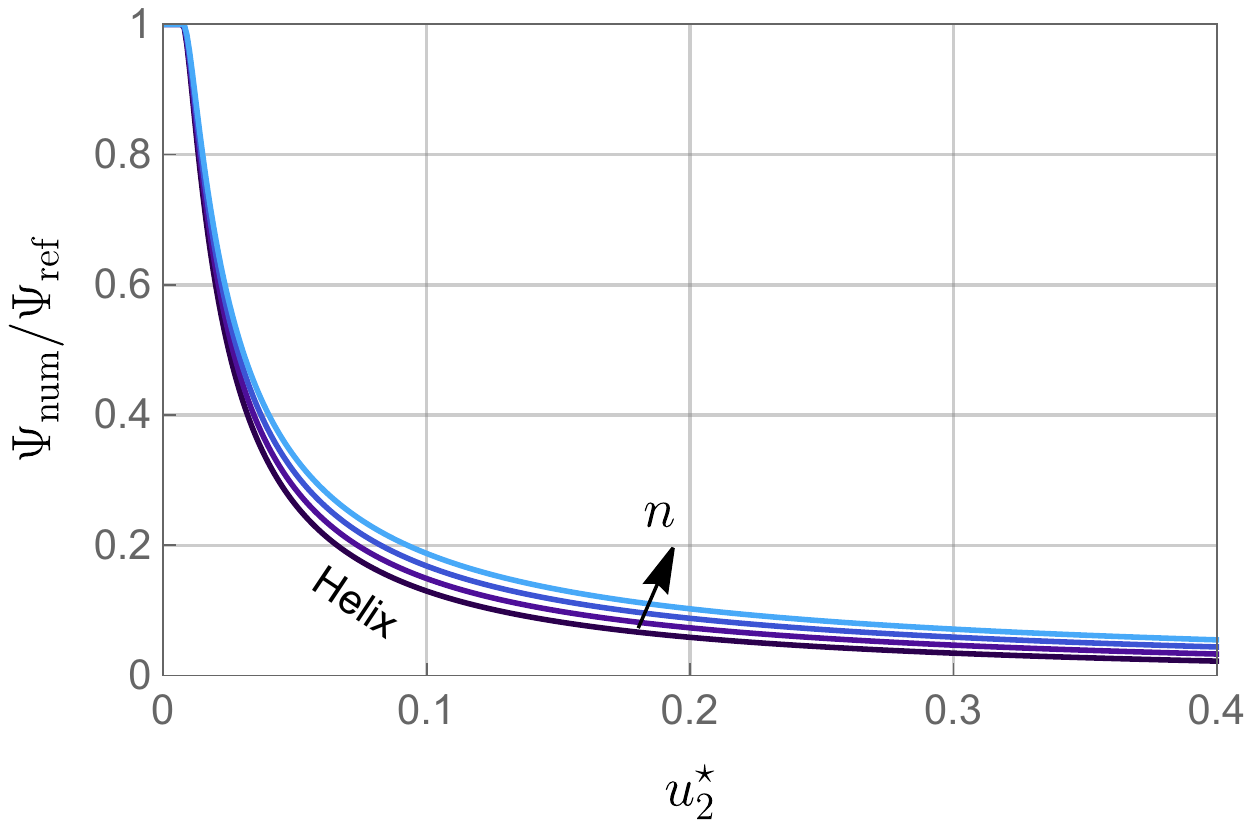}%
\includegraphics[width=0.5\textwidth]{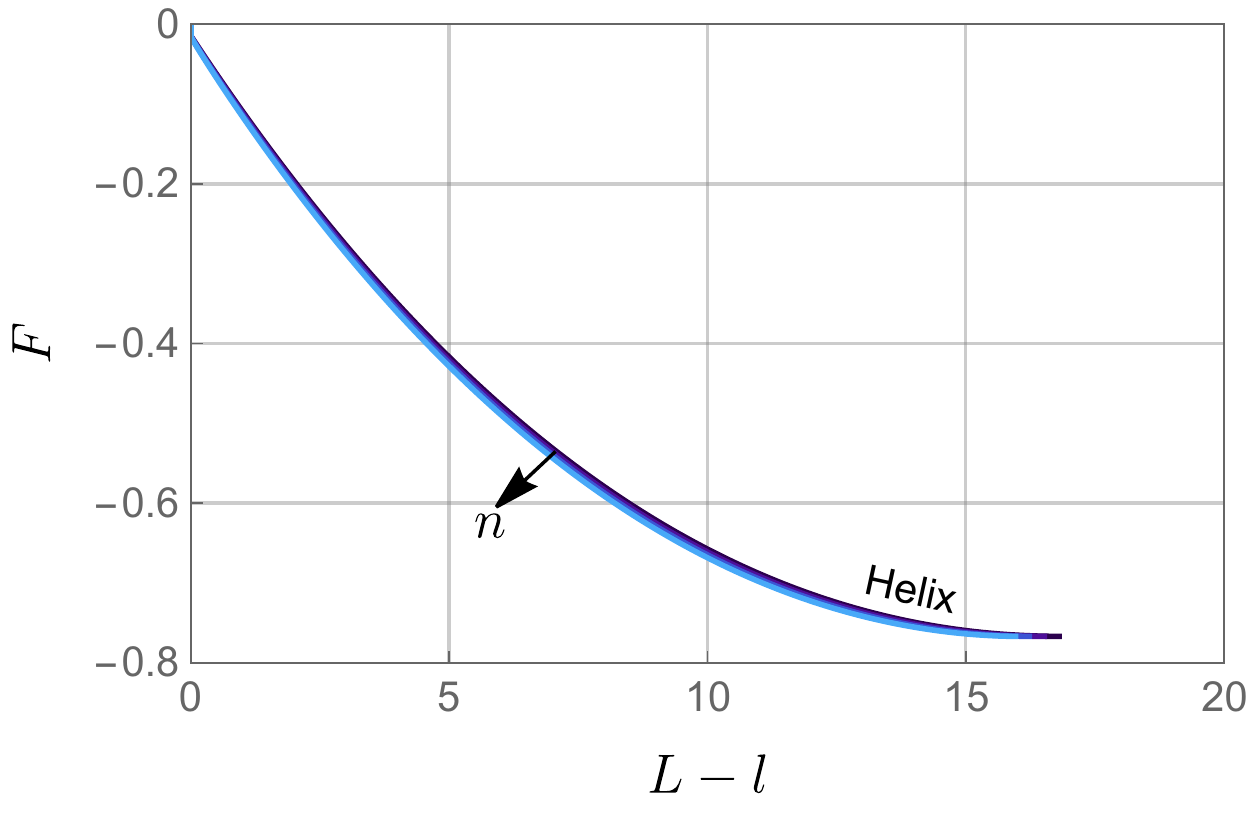}
\caption{Ratio between the energy of the buckled rod $\psi_\text{num}$ computed numerically and the theoretical energy of the straight unbuckled beam as a function of $u_2^\star$ (left) and force-displacement curves (right) for a rectangular cross section with $h/t=10$, $L=40$, $\chi=1$ and $\nu=0.35$. The curves corresponds to the helical and the modes exhibiting $n$ perversions (with $n=1,\,2,\,3$).}
\label{fig:energies}
\end{figure}
These results are analogous to what is observed in other similar systems. In fact, many elastic structures can exhibit helices and perversions as a result of buckling: examples include the case of an isolated rod with intrinsic curvature kept in tension by the action of an external force \cite{mcmillen2002tendril,Domokos_2005}, birod systems (assemblies of two rods clamped together by rigid connectors) \cite{Lessinnes_2016}, and elastic bilayers with pre-strain (where the layers are stretched before being glued together) \cite{Gerbode_2012,Huang_2012,Liu_2016,Lestringant_2017,DeSimone_2017,Caruso_2018}.

Despite the extensive literature on perversions, it is still unclear why shapes with multiple perversions are experimentally observable, even though they do not correspond to the critical modes as predicted by a linear stability analysis. In \cite{Lestringant_2017}, Lestringant and Audoly propose an explanation for the emergence of multiple perversions in an elastic bilayer. According to their study, the selection of the critical wavelength is dictated by the aspect ratio of the bilayer and the applied pre-strain. In their analysis, the deformability of the cross-section is of fundamental importance to achieve this conclusion. However, for the case of a single slender rod, the assumption of a non-deformable cross-section seems reasonable.

An alternative explanation for the appearance of multiple perversions may be related to the energy landscape of these mechanical systems. As we have shown in section \ref{sec:linear_analysis}, all the bifurcation points are very close for large $L$, see \eqref{eq:F_cr}-\eqref{eq:sol_pin_end}. Since the marginal stability thresholds are very similar, we argue that the share of elastic energy of the perversion is negligible with respect to the total energy of the system.
The results of our numerical simulations support this hypothesis: in figure~\ref{fig:energies} (left), 
we plot the ratio between the energy of the buckled and of the straight configuration as a function of the control parameter $u_2^\star$. In particular, the curves are related to different bifurcated branches, corresponding to a helical configuration or shapes exhibiting $n$ perversions. Remarkably, all the curves are very close one to another. A similar result is obtained when an axial force $F$ is applied: the force-displacement equilibrium curves for all the buckled configurations are very close one to another, see figure~\ref{fig:energies}~(right), and hence also the respective elastic energies.
This implies that the energy landscape of the system for fixed values of the control parameters is nearly flat, so that the non critical buckling modes may be energetically favorable in the presence of imperfections. This seems in agreement with the conjecture in \cite{Wang_2020}, where the authors state that the experimental observation of multiple perversions is caused by small perturbations and imperfections. We believe that, in our experiments, the friction between the wire and the connectors is responsible for the multi-stable behavior of the rods.

\section{Concluding remarks}
\label{sec:conclusions}

In this work, we have constructed and analyzed a model of an elastic beam free to slide along a rigid constraint and subjected to an external axial force. The assumption of large displacements  coupled with the possibility of the rod to slide and rotate about a rigid curve $\vect{w}$ gives rise to a highly nonlinear system, exhibiting multiple equilibrium configurations depending on the natural curvatures and twist of the beam.
Compared with the problem of a single rod deforming in space \cite{mcmillen2002tendril}, the introduction of the rigid constraint generates additional difficulties but, remarkably, allows us to write the total energy of the system as a functional depending on a single unknown, namely the angle $\omega$ between the normal to the curve $\vect{w}$ and the director $\vect{d}_1$.

The presence of a straight support significantly increases the critical load of the beam as compared with the classical Euler's buckling problem: in fact, for helical configurations, the buckling load \eqref{eq:dimensional_buckling_load} does not depend on the rod length and has a finite value also in the limit case $L\rightarrow +\infty$, contrarily with Euler's formula \eqref{eq:Euler}. In the pinned ends case, the helical mode is suppressed and thus perversions appear, according to the bifurcation thresholds \eqref{eq:dimensional_buckling_load2}, which are given by the buckling load for helices plus a term proportional to $L^{-2}$. Interestingly, in the limit of an infinite beam, all the bifurcation points collapse and the buckling load for the helices and for the configurations exhibiting perversions coincide.

Our analysis allows us to unfold and explore the nonlinear behavior of this system. The nearly flat energy landscape appears to be the main cause of the experimentally observed multi-stability of the beams: the energy necessary to generate a perversion is negligible with respect to the total energy of the rod whenever $L\gg c$. We argue that these features of the energy functional might be shared also by similar systems, such as a single rod with intrinsic curvature subjected to a tensile force \cite{mcmillen2002tendril}, and be the cause of the observation of multiple perversions in both artificial and natural structures, such as plant tendrils \cite{mcmillen2002tendril,Gerbode_2012}. These results may lead to applications in the design of shape morphing \cite{Huang_2012,Liu_2016,Su_2020} and deployable structures \cite{Pellegrino_2001,Lachenal_2012,noselli2019smart}. In particular, the possibility of controlling the force-displacement curves for helices by controlling the natural curvature $u_2^\star$, as shown in figure~\ref{fig:force_disp}, can be exploited for the design of compliant devices with tunable stiffness.

Future efforts will be devoted to the study of this system accounting for the friction between the constraint and the beam. This aspect is relevant for applications and is left as a possible future investigation. Moreover, we will enrich the analysis of the proposed model by considering the case in which $\vect{w}$ is not a rigid curve but a rod by itself. This last case has important implications in the modeling of more complicated assemblies of rods inspired by the motility of micro-organisms \cite{arroyo2014shape,noselli2019smart,riccobelli2020mechanics}.

\subsubsection*{Data Accessibility} The source code used for the numerical simulations is available on GitHub: \url{https:
//github.com/riccobelli/coiling-rod}.
\subsubsection*{Authors' Contributions} All the authors contributed to conceive and design the study. GN and DR
constructed the mathematical model and performed the experiments. DR performed the stability analysis and
the numerical simulations. ADS supervised research. All the authors drafted and approved the manuscript
for publication and agree to be held accountable for the work performed therein.
\subsubsection*{Competing Interests} We declare we have no competing interests.
\subsubsection*{Funding} The authors acknowledge the support of the European Research Council (AdG-340685-
MicroMotility).
\subsubsection*{Acknowledgements} The authors are members of the Gruppo Nazionale di Fisica Matematica -- INdAM

\bibliographystyle{abbrv}
\bibliography{refs}
\end{document}